\documentclass[11pt,a4paper,final]{article}
\usepackage{amsmath}
\usepackage{amsfonts}
\usepackage{amssymb}
\usepackage{amsthm}
\usepackage{mathrsfs}
\usepackage{mathtools}
\usepackage{dsfont}
\usepackage{graphicx}
\usepackage{color}
\usepackage[round]{natbib}
\usepackage[margin=0.9in]{geometry}
\usepackage[titletoc]{appendix}
\usepackage{bm}
\usepackage{algpseudocode}
\usepackage{algpascal}
\usepackage{afterpage}
\usepackage{enumitem}
\usepackage{multirow}
\usepackage{booktabs}

\usepackage{graphicx}
\usepackage{graphics}
\usepackage{pict2e}
\usepackage{epic}

\usepackage{epstopdf} 
\usepackage[ruled,vlined]{algorithm2e}
\usepackage{float}
\usepackage{tikz}
\usepackage{tkz-tab}
\usepackage[font = small]{caption}
\usepackage{subcaption}
\usepackage{latexsym}
\usetikzlibrary{shapes, decorations, bayesnet}

\usepackage[backref=page]{hyperref}
\hypersetup{
    colorlinks=true,
    citecolor=blue,
    linkcolor=red,
    filecolor=blue,      
    urlcolor=blue,
    pdfpagemode=FullScreen
    }

\newlength{\vslength}
\newtheorem{theorem}{Theorem}[section]

\newtheorem{lemma}{Lemma}[section]

\newtheorem{proposition}{Proposition}[section]

\newtheorem{corollary}{Corollary}[section]

\theoremstyle{remark}

\newtheorem{example}{\bf Example}[section]

\renewenvironment{proof}{\noindent{\it{Proof.} }}{\qed}

\newcommand{\bbR}{{\mathbb R}}

\newcommand{\bbN}{{\mathbb N}}

\newcommand{\bbeta}{\bm{\beta}}

\newcommand{\btheta}{\bm{\theta}}

\newcommand{\bpi}{\bm{\pi}}

\newcommand{\ie}{{\it i.e. }}

\newcommand{\bc}{\begin{center}}
\newcommand{\ec}{\end{center}}
\newcommand{\be}{\begin{equation}}
\newcommand{\ee}{\end{equation}}
\newcommand{\ba}{\begin{array}}
\newcommand{\ea}{\end{array}}
\newcommand{\bean}{\setlength\arraycolsep{1pt}\begin{eqnarray*}}
\newcommand{\eean}{\end{eqnarray*}}
\newcommand{\bea}{\setlength\arraycolsep{1pt}\begin{eqnarray}}
\newcommand{\eea}{\end{eqnarray}}
\newcommand{\ben}{\begin{enumerate}}
\newcommand{\een}{\end{enumerate}}
\newcommand{\bed}{\begin{itemize}}
\newcommand{\eed}{\end{itemize}}

\def\log{\text{log }}

\def\boxit#1{\vbox{\hrule\hbox{\vrule\kern6pt
 \vbox{\kern6pt#1\kern6pt}\kern6pt\vrule}\hrule}}

\def\bse{\begin{eqnarray*}}
\def\ese{\end{eqnarray*}}
\def\be{\begin{eqnarray}}
\def\ee{\end{eqnarray}}
\def\bq{\begin{equation}}
\def\eq{\end{equation}}
\def\bse{\begin{eqnarray*}}
\def\ese{\end{eqnarray*}}

\def\boxit#1{\vbox{\hrule\hbox{\vrule\kern6pt
          \vbox{\kern6pt#1\kern6pt}\kern6pt\vrule}\hrule}}

\def\log{\text{log }}

\def\bse{\begin{eqnarray*}}
\def\ese{\end{eqnarray*}}
\def\be{\begin{eqnarray}}
\def\ee{\end{eqnarray}}
\def\bq{\begin{equation}}
\def\eq{\end{equation}}
\def\bse{\begin{eqnarray*}}
\def\ese{\end{eqnarray*}}

\newcommand{\uk}       {\boldsymbol{k}}

\newcommand{\un}       {\mbox{\boldmath$n$}}

\newcommand{\ut}       {\boldsymbol{t}}

\newcommand{\uu}       {\boldsymbol{u}}

\newcommand{\uX}       {\boldsymbol{X}}
\newcommand{\ux}       {\boldsymbol{x}}
\newcommand{\uY}       {\boldsymbol{Y}}

\newcommand{\uz}       {\boldsymbol{z}}

\newcommand{\ubeta}             {\mbox{\boldmath$\beta$}}

\newcommand{\uiota}             {\mbox{\boldmath$\uiota$}}

\newcommand{\upi}               {\mbox{\boldmath$\pi$}}

\newcommand{\uphi}              {\mbox{\boldmath$\phi$}}

\numberwithin{equation}{section}
\usepackage{comment}

\graphicspath{{Figures/}}

\begin{document}
\thispagestyle{empty}
\title{
    \vspace*{-9mm}
    Blocked Gibbs sampler for hierarchical Dirichlet processes
}
\author{Snigdha Das$^{1}$, Yabo Niu$^{2}$, Yang Ni$^{1}$, Bani K. Mallick$^{1}$, Debdeep Pati$^{1}$ \\[2mm]
    {\small\it {}$^{1}$Department of Statistics, Texas A$\&$M University} \\
    {\small\it {}$^{2}$Department of Mathematics, University of Houston}
}
\date{\today}
\maketitle

\vspace*{-10mm}
\begin{abstract}
Posterior computation in hierarchical Dirichlet process (HDP) mixture models is an active area of research in nonparametric Bayes inference of grouped data.  Existing literature almost exclusively focuses on the Chinese restaurant franchise (CRF) analogy of the marginal distribution of the parameters, which can mix poorly and has a quadratic complexity with the sample size. 
A recently developed slice sampler allows for efficient blocked updates of the parameters, but is shown to be statistically unstable in our article. We develop a blocked Gibbs sampler that employs a truncated approximation of the underlying random measures to sample from the posterior distribution of HDP, which produces statistically stable results, is highly scalable with respect to sample size, and is shown to have good mixing. The heart of the construction is to endow the shared concentration parameter with an appropriately chosen gamma prior that allows us to break the dependence of the shared mixing proportions and permits independent updates of certain log-concave random variables in a block.  {\em En route}, we develop an efficient rejection sampler for these random variables leveraging piece-wise tangent-line approximations.
    \medskip
    
\end{abstract}

\section{Introduction}
Hierarchical Dirichlet process (HDP) \citep{Teh2006HierarchicalDP} is a widely popular Bayesian nonparametric approach towards model-based clustering of grouped data, where each group is characterized by a mixture model and mixture components are shared between groups. It finds a variety of applications in statistical and machine learning tasks such as information retrieval \citep{CowansIR2004}, topic modeling \citep{Teh2006HierarchicalDP}, multi-domain learning \citep{Canini2010}, multi-population haplotype phasing \citep{Sohn2008AHD}, 
to name a few. It also serves as a prior for hidden Markov models \citep{Teh2006HierarchicalDP} with applications in speaker diarization \citep{Fox2011AAS}, word segmentation \citep{goldwater-etal-2006-contextual}, among many others. 
The process has been studied from an analytical perspective using hierarchical normalized completely random measures \citep{camerlenghi2017bayesian, camerlenghi2019distribution, Argiento2020HierarchicalNC}, along with extensions in several contexts, including but not limited to species sampling models \citep{bassetti2020speciessampling}, survival analysis using dependent mixture hazard rates \citep{Camerlenghi2021SurvivalAV}.

 The widespread popularity of HDP has motivated several sampling-based Markov chain Monte Carlo (MCMC) algorithms \citep{Teh2006HierarchicalDP, Fox2011AAS, Amini2019ExactSS}, as well as variational approximations \citep{Teh2007VI, Wang2011VI, NIPS2012_Bryant}, for efficient posterior inference. Split-merge MCMC \citep{WangBlei2012, NIPS2014ChangFisher}, parallel MCMC \citep{Asuncion2008, Williamson13ICML, cheng2014parallel}, and mini-batch online Gibbs \citep{Kim_OnlineGibbs2016} algorithms  were also proposed. We focus on sampling-based approaches in this paper, 
 the most prominent implementation of which is the Chinese restaurant franchise (CRF) based collapsed Gibbs sampler \citep{Teh2006HierarchicalDP}, that marginalizes over the random probability measures of HDP.  The conditional distributions of the global and local cluster assignments of each data point depends on the assignments of all the other data points, incorporating auto-correlation between the parameters and mandating the need to iterate over each observation for each group. These one-at-a-time updates suffer when the sample size is large, motivating the need for a blocked Gibbs sampler, analogous to that of a Dirichlet process (DP) \citep{Ishwaran2001GibbsSM}, that is scalable and exhibits good mixing behavior. A straightforward formulation is computationally challenging due to the lack of conjugacy in the posterior updates of the global mixture weights lying on a simplex, whose dimension can be potentially large as they are shared by all the groups. To address this, a recent unpublished paper \citep{Amini2019ExactSS} proposed an exact slice sampler that introduces latent variables to allow for natural truncation of the underlying infinite dimensional Dirichlet measures and conjugate blocked updates of the global weights, which however incurs some instability in posterior inference, as demonstrated in our simulations.

In this paper, we develop a blocked Gibbs sampler for HDP by considering a finite-dimensional truncation of the underlying DPs as in \eqref{finite_HDP}. 
In addition, the concentration parameter shared by the local DPs is endowed with a gamma prior whose shape parameter is chosen as the concentration of the global DP.
The algorithm (i) identifies the global weights as a normalization of certain log-concave random variables by marginalizing over the shared concentration, that renders an equivalent prior specification in terms of the unnormalized weights which are dependent in their non-standard posterior, (ii) introduces auxiliary gamma random variables as a slice, which grants conditional independence to the posterior of the unnormalized weights in the form of a tilted gamma density, and (iii) devises a careful envelope for this density to perform an exact rejection sampler with very high acceptance probability.  A critical difference of our blocked Gibbs sampler with the slice sampler of \cite{Amini2019ExactSS} is to avoid using the atoms of local DPs as latent variables. This resulted in a non-standard full conditional distribution of the global weight variables, but leads to better statistical inference compared to \citet{Amini2019ExactSS}. 


Our algorithm exhibits good mixing behavior with better effective sample size of distinct atoms drawn apriori from the base measure. The computation time is stable with the sample size, since the blocked parameter updates factorize across observations in each group that allows for parallelization and makes it suitable for applications to large data sets. Additionally, it refrains from the involved bookkeeping of the CRF metaphor, enabling a simpler implementation of HDP for practitioners.

\section{Framework}
\subsection{The HDP mixture model}
HDP \citep{Teh2006HierarchicalDP} defines a nonparametric prior over parameters for grouped data.
Using $j$ to index $J$ groups of data and $i$ to index observations within a group, let $\ux_j = (x_{j1}, x_{j2},\ldots, x_{jn_j})$ denote $n_j$ observations in group $j$ and $\theta_{ji}$ denote the parameter specifying the mixture component associated with $x_{ji}$, assuming each $x_{ji}$ is drawn independently from a mixture model. For each $j$ and $i$, the mixture model is given as $\theta_{ji} \mid G_j \sim G_j$, $x_{ji} \mid \theta_{ji} \sim F(\theta_{ji}) $,
where $G_j$ denotes a prior distribution for $\btheta_j = (\theta_{j1}, \theta_{j2}, \ldots)$ and $F(\theta_{ji})$ denotes the distribution of $x_{ji}$ given $\theta_{ji}$. 
Let $\text{DP}(\alpha, G)$ denote a Dirichlet process with base measure $G$ and concentration parameter $\alpha$. 
HDP defines a global random measure $G_0 \sim \text{DP}(\gamma, H)$ and a set of random probability measures $G_j \sim \text{DP}(\alpha_0, G_0)$, which are conditionally independent given $G_0$. The baseline probability measure $H$ dictates the prior distribution for the factors $\theta_{ji}$. The global and shared concentration parameters $\gamma$ and $\alpha_0$ govern the variability by which $G_0$ varies around $H$ and $G_j$ varies around $G_0$ for each $j$, respectively.

The process has been discussed using several representations. The stick-breaking construction of HDP allows us to express $G_0$ and $G_j$ respectively as $G_0 = \sum_{k=1}^\infty \beta_k \delta_{\phi_k}$ and $G_j =  \sum_{k=1}^\infty \pi_{jk} \delta_{\phi_k}$,
where $\phi_k \sim H$ independently and $\ubeta = (\beta_k)_{k=1}^\infty \sim \text{GEM}(\gamma)$ are mutually independent; and $\upi_j = (\pi_{jk})_{k=1}^\infty \sim \text{DP}(\alpha_0, \ubeta)$. GEM stands for Griffiths, Engen and McCloskey; see \cite{pitman2002poissondirichlet} for details. The marginal probabilities obtained from integrating over the random measures $G_0$ and $G_j$ are described in terms of the Chinese restaurant franchise metaphor. The CRF consists of $J$ restaurants, each having an infinite number of tables and sharing an infinite global menu of dishes ($\phi_k$). Each restaurant corresponds to a group and each $\theta_{ji}$ corresponds to a customer seated at table $t_{ji}$. A dish $k_{jt}$ at table $t$ of restaurant $j$,  is served from the global menu and is shared by all customers seated at that table. The table indices $t_{ji}$ and dish indices $z_{ji} \ (= k_{jt_{ji}})$ respectively correspond to the local and global clustering labels.
Further, an HDP mixture model is derived as the infinite limit of the following collection of finite mixture models,
\begin{align} 
\label{finite_HDP}
    \nonumber \ubeta \mid \gamma & \sim \text{Dir}\left({\gamma}/{L}, \ldots, {\gamma}/{L}\right)
    &\\
     \upi_j \mid \beta, \alpha_0 & \sim \text{Dir}\left(\alpha_0 \ubeta \right)       
    &  z_{ji} \mid \upi_j &\sim \upi_j\\
    \nonumber \phi_k \mid H &\sim H  
    &  x_{ji} \mid z_{ji}, \{\phi_k \}_{k=1}^L &\sim F(\phi_{z_{ji}})
\end{align}
where $\ubeta = (\beta_1, \ldots, \beta_L)$ and $\upi_j = (\pi_{j1}, \ldots, \pi_{jL})$ denote the global and group-specific mixing weights, and $z_{ji}$ denotes the mixture component associated with $x_{ji}$. 
Following \cite{Ishwaran2002ExactAA}, \cite{Teh2006HierarchicalDP} proved that as $L \rightarrow \infty$, the random probability measures, $G_0^L = \sum_{k=1}^L \beta_k \delta_{\phi_k}$ and $G_j^L = \sum_{k=1}^L \pi_{jk} \delta_{\phi_k}$ converge to $G_0$ and $G_j$ respectively, and hence the marginal distribution induced on $\ux$ by this finite model approaches the HDP mixture model.

\subsection{Overview of existing posterior sampling algorithms}
\label{overview_algorithms}
Three Gibbs samplers \citep{Teh2006HierarchicalDP} were initially proposed for posterior inference on the HDP mixture model. The first is a collapsed sampler that marginalizes out $G_0$ and $G_j$, and updates $\ut = (t_{ji})_{j,i}$ and $\uk = (k_{jt})_{j, t}$ sequentially. The second samples $\ubeta = (\beta_k)_k$ using an augmented scheme and updates the indices $\ut_j = (t_{ji})_i$ and $\uk_j = (k_{jt})_t$ for each group separately. The third uses a direct assignment scheme to sample $\uz = (z_{ji})_{j,i}$, along with sampling $\ubeta$ and $\ut$.  
The algorithms involve heavy bookkeeping  and iterate over each observation for every group, since the update of a cluster label for each data point is conditioned on all other data points.  This induces high auto-correlation between the parameters and issues with mixing of the Markov chain, in addition to slow scaling. \cite{Fox2011AAS} built upon the hierarchical Dirichlet process hidden Markov model (HDP-HMM) \citep{Teh2006HierarchicalDP} to introduce a sticky HDP-HMM and gave a blocked Gibbs sampler using the truncated model in \eqref{finite_HDP}, which updates auxiliary variables (motivated by the CRF metaphor) by iterating over observations in each group and uses them to give blocked updates of parameters in HDP-HMM. \cite{WangBlei2012} proposed a split-merge MCMC algorithm, followed by its parallel implementation by \cite{NIPS2014ChangFisher}, who argued that the former showed a marginal improvement over the CRF based samplers. To boost computation time and memory, parallel MCMC algorithms that split the data across multiple processors were proposed. 
While \cite{Asuncion2008} used approximations to propose a parallel Gibbs sampler for HDP, \cite{Williamson13ICML} used the fact that Dirichlet mixtures of DPs are DPs and introduced auxiliary variables in terms of process indicators to obtain conditionally independent local sampling of different HDPs (using the CRF scheme) across each processor, when the concentration parameter of the bottom level DPs has a gamma prior. The process indicators are updated globally using a Metropolis Hastings (MH) step to intermittently move data across clusters. 
Although exact parallel MCMC methods have gained popularity, \cite{Gal14-PMLR-v32} showed that the algorithm of \cite{Williamson13ICML} results in an unbalanced distribution of data to different processors due to the sparseness properties of the Dirichlet distribution used for re-parametrisation and the exponential decay in the size of clusters in a DP. Consequently, even if a large number of processors is available, only a small subset of it would be used in practice. 
As an improvement over this unbalanced workload, \cite{cheng2014parallel} proposed a parallel Gibbs sampler for HDP by exploring the equivalence between its generative process and that of the gamma-gamma-Poisson process (G2PP). The algorithm relies on the augmented G2PP model and reconstructs the dataset by combining bootstrap and reversible jump MCMC techniques to enable independent updates of mixture components.
\cite{Kim_OnlineGibbs2016} derived a mini-batch online Gibbs sampler for HDP as an alternative to online variational algorithms, by proposing a generalized HDP prior that approximates the posterior of HDP parameters in each batch of data.
More recently, \cite{Amini2019ExactSS} proposed an exact slice sampler for HDP, by introducing atoms of the bottom level DPs as latent variables, that allows for natural truncation of $G_0$ and $G_j$ and conjugate blocked updates of $\ubeta$. Although the algorithm mixes fast and is suited for parallel implementation, it comes at the cost of some instability in statistical accuracy of posterior estimates, as we demonstrate in \S\ref{simulation}.

\subsection{Construction of a blocked Gibbs sampler} 
\label{blockedGibbs}

For DP mixture models, it is well known that blocked Gibbs sampling \citep{Ishwaran2001GibbsSM} has significant improvements in terms of both mixing and scalability over the Chinese restaurant process (CRP) based collapsed samplers. The blocked Gibbs algorithm replaces the infinite dimensional DP prior by its finite dimensional approximation \citep{Ishwaran2002ExactAA} allowing the model parameters to be expressed entirely in terms of a finite number of random variables, which are updated in blocks from simple multivariate distributions. 
Instead of adapting the CRF-based samplers to make them efficient and scalable, we focus on devising a similar blocked Gibbs sampler for HDP in this paper, which will adhere to both improved mixing and scaling over large sample sizes. To that end, we use the finite approximation in \eqref{finite_HDP} and require the shared concentration parameter $\alpha_0$ to have a gamma prior. We shall discuss the importance of this specification in \S\ref{samp_beta}, where we describe the sampling scheme for the global weights $\ubeta$.

The joint posterior of the parameters in \eqref{finite_HDP} with $\alpha_0 \sim \text{Gamma}(a_0, b_0)$, is as follows
\begin{align} 
\label{jointposterior}
    \nonumber p(\uphi, \uz, \upi, \ubeta, \alpha_0 \mid \ux) \propto 
    & \ p(\ux \mid \uz, \uphi) \ p(\uz \mid \upi) \ p(\uphi) \ p(\upi \mid \ubeta, \alpha_0) \ p(\ubeta) \ p(\alpha_0)\\
    \nonumber \propto & \prod_{j=1}^J \prod_{i = 1}^{n_j} \prod_{k=1}^L \left\{\pi_{jk} f(x_{ji} \mid \phi_k)\right\}^{\mathds{1}_{\{z_{ji} = k\}}} \prod_{k=1}^L h(\phi_k) \ \prod_{k=1}^L \beta_k^{\frac{\gamma}{L} - 1}\ \mathds{1}_{\left\{ \ubeta \in \mathcal{S}^{L-1} \right\}} \ \times \\
    & \prod_{j=1}^J \frac{\Gamma(\alpha_0)}{\prod_{k=1}^L\Gamma(\alpha_0 \beta_k)} \prod_{k=1}^L \pi_{jk}^{\alpha_0 \beta_k - 1} \ \mathds{1}_{\left\{\upi_j \in \mathcal{S}^{L-1} \right\}} \ \alpha_0^{a_0 - 1 }e^{-b_0 \alpha_0}  \mathds{1}_{\left\{ \alpha_0 > 0 \right\}}
\end{align}
where $\mathcal{S}^{L-1}$ denotes the $L$-dimensional simplex.
The blocked parameter updates are enlisted in \S\ref{full_cond} of the Supplement. A key aspect is that the posterior of $\uz$  factorizes across observations in each group, ensuring scalability to large sample sizes. Moreover, this approach refrains from the CRF metaphor and its heavy bookkeeping, allowing a simpler implementation of the HDP mixture model for practitioners. 

The full conditionals of the parameters reveal that $\uphi$ (if $H$ is chosen conjugate to $F$), $\uz$ and $\upi$ have closed form conjugate updates while the posteriors of $\ubeta$ and $\alpha_0$ have non-standard forms. $\alpha_0$ being a one-dimensional variable, can be easily sampled from its posterior using an MH algorithm. The main computational bottleneck is the non-standard density of the global weights $\ubeta$, which lie on a simplex, as follows
\begin{equation*}
        p(\ubeta \mid \ux, \uphi, \uz, \upi,\alpha_0 ) \propto \ \prod_{k=1}^L \ \frac{1}{ \Gamma(\alpha_0 \beta_k)^J} \ \big(\prod_{j=1}^J \pi_{jk}\big)^{\alpha_0 \beta_k}\, \beta_k^{\,\frac{\gamma}{L} -1}\ \mathds{1}_{\left\{ \ubeta \,\in \, \mathcal{S}^{L-1} \right\}}
\end{equation*}
The dimension $L$ of the global weights can be potentially large in practice since the distinct global atoms are shared by all the groups, which poses challenges to their sampling.

\section{Sampling of global weights using a slicing normalization} 
\label{samp_beta}

As a remedy to overcome the potential shortcoming of a blocked Gibbs sampler, we explore an equivalent representation of the hierarchical priors on the weights, $\upi$ and $\ubeta$, and the shared concentration parameter $\alpha_0$, when the shape parameter $a_0$ of its gamma prior is chosen as the concentration parameter $\gamma$ of the global DP. Such a specification enables $\ubeta$ to be expressed as normalization of non-negative random variables that can be conveniently sampled from its posterior using a suitable augmentation scheme.

When $a_0$ is chosen to be $\gamma$, the hierarchical structure of the priors on $\upi, \ubeta, \alpha_0$ is as follows,
\begin{align}
 \label{prior_pi_beta_alpha0}
    \bpi_j \mid \ubeta, \alpha_0 & \sim \text{Dir}\left(\alpha_0 \bbeta \right)       
    &  \ubeta & \sim \text{Dir}\left({\gamma}/{L}, \ldots, {\gamma}/{L}\right)  
    & \alpha_0 \sim \text{Gamma}(\gamma, b_0)
\end{align}
It is well known that a set of gamma-distributed variables normalized by their sum is Dirichlet distributed. Additionally,  their sum, characterized by a gamma distribution, remains independent of the resulting Dirichlet distribution.
Thereby, considering a vector of non-negative random variables $\ut = (t_1, t_2, \ldots, t_L)$ such that $t_k \sim \text{Gamma}(\gamma/L, b_0)$ independently for each $k$, (\ref{prior_pi_beta_alpha0}) can be an equivalently represented by
\begin{align}
\label{prior_pi_T}
    \bpi_j \mid\ut & \sim \text{Dir}\left(\ut\right) & t_k & \sim \text{Gamma}\left(\gamma/L, b_0\right)
 \end{align}
where $\ubeta$ is viewed as a normalized vector of $\ut$ and $\alpha_0$ denotes the sum of all elements of $\ut$.

This hierarchical specification can alternatively be formulated by marginalizing out $\alpha_0$ having a Gamma$(a_0, b_0)$ prior, wherein the joint prior on $\upi$ and $\ubeta$ boils down to 
\begin{equation}
\label{pi_beta}
    p(\upi, \ubeta) \propto \int_{0}^\infty  t ^{L - \gamma + a_0-1} \ \Gamma(t)^J \prod_{k=1}^L \frac{e^{-b_0 t \beta_k} (t \beta_k)^{\,\frac{\gamma}{L} -1}}{ \Gamma(t \beta_k)^J} \prod_{j=1}^J \pi_{jk}^{t\beta_k-1} \ \mathds{1}_{\left\{\upi_j \, \in \, \mathcal{S}^{L-1}\right\}}  \mathds{1}_{\left\{\ubeta \, \in \, \mathcal{S}^{L-1}\right\}}\ dt
\end{equation}
When $a_0$ is chosen to be $\gamma$ in \eqref{pi_beta}, using Lemma \ref{lemma:normalization} in \S{\ref{lemmas}} of the Supplement, $\ubeta$ can be observed to be a normalized vector of random variables $\ut = (t_1, t_2, \ldots,t_L)$ whose joint density with $\upi$ is given by
\begin{equation*}
    p(\upi,\ut) \propto \Gamma \bigg(\sum_{k=1}^L t_k \bigg)^J \prod_{k=1}^L \frac{e^{-b_0 t_k} \ t_k^{\,\frac{\gamma}{L} -1}}{ \Gamma(t_k)^J} \prod_{j=1}^J \pi_{jk}^{t_k-1} \ \mathds{1}_{\left\{\upi_j \, \in \, \mathcal{S}^{L-1}\right\}} \ \mathds{1}_{\left\{ t_k > 0 \right\}},
\end{equation*}
and is equivalent to the hierarchical specification in \eqref{prior_pi_T}.


This specification is computationally convenient since posterior samples of $\ubeta$ can by obtained by sampling $\ut$ from its corresponding posterior and setting $\beta_k = t_k \big/ \sum_{l = 1}^L t_l$ for each $k$. More specifically, samples from $p(\uphi, \uz, \upi,\ubeta \mid \ux)$ by collapsing out $\alpha_0$ can be obtained by sampling from the augmented posterior $p(\uphi, \uz, \upi,\ut \mid \ux)$. 
Writing the augmented posterior explicitly
\begin{align} 
\label{aug_post_1}
    \nonumber p(\uphi, \uz, \upi,\ut \mid \ux) \propto \
    &  p(\ux \mid \uz, \uphi) \ p(\uz \mid \upi) \ p(\uphi) \ p(\upi \mid\ut) \ p(\ut)\\
    \nonumber \propto \ & \prod_{j=1}^J \prod_{i = 1}^{n_j} \prod_{k=1}^L \left\{\pi_{jk} f(x_{ji} \mid \phi_k)\right\}^{\mathds{1}_{\{z_{ji} = k\}}} \prod_{k=1}^L h(\phi_k) \ \times \\
    & \ \Gamma \bigg(\sum_{k=1}^L t_k \bigg)^J \prod_{k=1}^L \frac{e^{-b_0 t_k} \ t_k^{\,\frac{\gamma}{L} -1}}{ \Gamma(t_k)^J} \prod_{j=1}^J \pi_{jk}^{t_k-1} \ \mathds{1}_{\left\{\upi_j \, \in \, \mathcal{S}^{L-1}\right\}} \ \mathds{1}_{\left\{ t_k > 0 \right\}}\,,
\end{align}
we observe that $\uphi$, $\uz$, and $\upi$ retain conjugacy in their blocked posterior updates, while the independence and conjugacy of $\ut$ is not preserved in its non-standard posterior. We eliminate this dependence by incorporating suitably chosen auxiliary variables, a technique widely employed in Bayesian nonparametrics involving normalized completely random measures. Noting that the term $\Gamma \big(\sum_{k=1}^L t_k \big)^J$ prevents the posterior density of $\ut$ from factorizing over $k$, we write it as 
\begin{equation*}
    \Gamma \bigg( \sum_{k=1}^L t_k \bigg)^J = \prod_{j=1}^J \left\{\int_0^\infty e^{-u_j}\,u_j^{\left(\sum_{k=1}^L t_k \right)-1} \, du_j \right\},
\end{equation*}
and introduce a slice in (\ref{aug_post_1}) using auxiliary random variables $\uu = (u_1, u_2, \ldots, u_J)$ to get the following augmented posterior of $\uphi$, $\uz$, $\upi$, $\ut$ and $\uu$,
\begin{align} 
\label{aug_post_final}
    \nonumber p(\uphi, \uz, \upi,\ut,\uu \mid \ux) \propto 
     & \prod_{j=1}^J \prod_{i = 1}^{n_j} \prod_{k=1}^L \left\{\pi_{jk} f(x_{ji} \mid \phi_k)\right\}^{\mathds{1}_{\{z_{ji} = k\}}} \prod_{k=1}^L h(\phi_k) \prod_{j=1}^J e^{-u_j}\,u_j^{\left(\sum_{k=1}^L t_k \right)-1} \mathds{1}_{\{u_j > 0\}} \, \times \\
    & \prod_{k=1}^L \frac{e^{-b_0 t_k} \ t_k^{\,\frac{\gamma}{L} -1}}{ \Gamma(t_k)^J} \prod_{j=1}^J \pi_{jk}^{t_k-1} \ \mathds{1}_{\left\{\upi_j \, \in \, \mathcal{S}^{L-1}\right\}} \ \mathds{1}_{\left\{ t_k > 0 \right\}}
\end{align}

\noindent The full conditional posteriors of $\uu$ and $\ut$ are then given by, 
\begin{align*}
    p(\uu \mid \ux, \uphi, \uz, \upi,\ut) & \propto \prod_{j=1}^J e^{-u_j}\,u_j^{\left(\sum_{k=1}^L t_k \right)-1} \ \mathds{1}_{\{u_j > 0\}}, & 
    p(\ut \mid \ux, \uphi, \uz, \upi,\uu) & = \prod_{k=1}^L f_k(t_k),
\end{align*}
where 
\begin{equation} 
\label{f_k}
    f_k(t) \propto \frac{1}{ \Gamma(t)^J} \ e^{-(b_0 - \sum_{j=1}^J \log \pi_{jk} - \sum_{j=1}^J \log  u_j) \, t }  \ t ^{\,\frac{\gamma}{L} -1} \ \mathds{1}_{\left\{ t > 0 \right\}}, \quad k = 1, 2, \ldots, L.
\end{equation}

\noindent Note that $\uu$ given $\ut$ is a vector of independent Gamma$\big( \sum_{k=1}^L t_k, 1 \big)$ random variables, which can be sampled easily, and $\ut$ given $\uu$ is a vector of independent random variables with density $f_k$.
We refer to $f_k$ as a tilted gamma density with parameters $J$, $A = \gamma/L$ and $B_k = b_0 - \sum_{j=1}^J \log  \pi_{jk} - \sum_{j=1}^J \log  u_j$.

\vspace{1ex}

\begin{figure}[htp]
     \centering
     \begin{subfigure}[b]{0.3\textwidth}
         \centering
         \resizebox{0.35\textwidth}{!}{%
	\begin{tikzpicture}
            \node[draw, shape = circle, inner sep = 4pt] at (0, 1.5) (beta) {\Large $\ubeta$};
            \node[draw, shape = circle, inner sep = 4pt] at (0, 0) (pi) {\Large $\upi$};
            \node[draw, shape = circle, inner sep = 3pt] at (-1.5, 0) (alpha) {\Large $\alpha_0$};
            \node[draw, shape = circle, inner sep = 4pt] at (0, -1.5) (z) {\Large $\uz$};
		\node[draw, regular polygon sides = 4, inner sep = 7pt] at (0, -3)(x) {\Large $\ux$};
            \node[draw, shape = circle, inner sep = 3pt] at (-1.5, -3) (phi) {\Large $\uphi$};
		
            \edge[->] {alpha} {pi};
		\edge[->] {beta} {pi};
            \edge[->] {pi} {z};
		\edge[->] {z} {x};
            \edge[->] {phi} {x};
	\end{tikzpicture}}
         \caption{}
     \end{subfigure}
     \begin{subfigure}[b]{0.3\textwidth}
         \centering
         \resizebox{0.45\textwidth}{!}{%
	\begin{tikzpicture}
            \node[draw, regular polygon sides = 4, inner sep = 7pt] at (0, 1.5)(x) {\Large $\ux$};
            \node[draw, shape = circle, inner sep = 4pt] at (1.1, 0) (z) {\Large $\uz$};
            \node[draw, shape = circle, inner sep = 3pt] at (-1.1, 0) (phi) {\Large $\uphi$};
            \node[draw, shape = circle, inner sep = 4pt] at (0, -1.5) (pi) {\Large $\upi$};
            \node[draw, shape = circle, inner sep = 3pt] at (-1.1, -3) (alpha) {\Large $\alpha_0$};
            \node[draw, shape = circle, inner sep = 3pt] at (1.1, -3) (beta) {\Large $\ubeta$};
		
            \edge[->] {x} {phi};
		\edge[->] {x} {z};
            \edge[<->] {phi} {z};
		\edge[<->] {z} {pi};
            \edge[<->] {pi} {alpha};
            \edge[<->] {pi} {beta};
            \edge[<->] {alpha} {beta};
	\end{tikzpicture}}
         \caption{}
     \end{subfigure}
     \begin{subfigure}[b]{0.3\textwidth}
         \centering
         \resizebox{0.61\textwidth}{!}{%
	\begin{tikzpicture}[implies/.style={thick, double equal sign distance,-implies},]
            \node[draw, regular polygon sides = 4, inner sep = 7pt] at (0, 1.5)(x) {\Large $\ux$};
            \node[draw, shape = circle, inner sep = 4pt] at (1.1, 0) (z) {\Large $\uz$};
            \node[draw, shape = circle, inner sep = 3pt] at (-1.1, 0) (phi) {\Large $\uphi$};
            \node[draw, shape = circle, inner sep = 4pt] at (0, -1.5) (pi) {\Large $\upi$};
            \node[draw, shade, shape = circle, inner sep = 4pt] at (0, -3.25) (t) {\Large $\ut$};
            \node[draw, shade, shape = circle, inner sep = 4pt] at (-1.8, -3.25) (u) {\Large $\uu$};
		\node[draw, shape = circle, inner sep = 3pt] at (1.7, -3.25) (beta) {\Large $\ubeta$};
  
            \edge[->] {x} {phi};
		\edge[->] {x} {z};
            \edge[<->] {phi} {z};
		\edge[<->] {z} {pi};
            \edge[<->] {pi} {t};
            \edge[implies] {t} {beta};
            \edge[<->] {t} {u};
	\end{tikzpicture}}
         \caption{}
     \end{subfigure}
    \vspace{-0.1 in}
    \caption{Graphical models showing the dependence of parameters in the (a) truncated HDP mixture model \eqref{finite_HDP}, (b) joint posterior \eqref{jointposterior}, and (c) augmented posterior (\ref{aug_post_final}). Shaded nodes and double arrows represent augmented variables and deterministic relations, respectively.}
    \label{fig:graphicalmodel}
\end{figure}

Our remaining task is to devise an algorithm to sample $t_k$ from its conditional density $f_k$, for each $k = 1, 2, \ldots L$. While an MH proposal is generally applicable for sampling from any non-standard density that is known upto constants, tuning a single MH proposal may not work in our setting due to the changing parameters of the target density with each Gibbs iteration. This suggests the need for more adaptive MH schemes to address the issue, which might not be very straightforward to implement.
Consequently, we develop an exact tuning-free rejection sampler to sample the $t_k$'s independently from $f_k$, exploiting the fact that $f_k$ in \eqref{f_k} a is log-concave density, as proved in Lemma \ref{lemma:log-concave} in \S\ref{lemmas} of the Supplement. We build a piece-wise tight upper envelope leading to a mixture of densities that can be conveniently sampled from, ensuring high acceptance rates.  The next subsection describes the construction of the envelope while deferring the sampling technique to \S\ref{samplerforbeta} of the Supplement.

\subsection{Construction of the cover for rejection sampling of the unnormalized global weights}
In the following, we describe the construction of the cover density, $g_k$ to get exact samples from our target density, $f_k$, $k = 1,2,\ldots, L$, using a rejection sampler, the general algorithm of which is given in \S\ref{rejectionsampler} of the Supplement. Let $\Tilde{f}_k(x) = \frac{1}{ \Gamma(x)^J} \ x ^{\, A -1} \ e^{-B_k x} \ \mathds{1}_{\left\{ x > 0 \right\}}$ and $C_{f_k} = \int_0^\infty \Tilde{f}_k(x) \, dx$ denote the density upto constants and its normalizing constant, respectively.
Note that $J \in \mathbb{N}$, $B_k \in \bbR$ and $A \in (0,1)$, since we are working with a finite truncation of an HDP mixture model by fixing the number of clusters ($L$) at a suitably chosen large number, as described in \S\ref{blockedGibbs}.
To ensure a high acceptance rate of the rejection sampler, we require a sharp upper bound for $\Tilde{f}_k$ that is easy to sample from.
This is achieved by using piece-wise tangent lines of the logarithm of $f_k$ at carefully chosen knot points, exploiting the concavity of the log density. 

We begin with noting that when the parameter $B_k$ is positive, for $x>0$, $x ^{\, A -1} \ e^{-B_k x} $ is decreasing and continuous, while $\Gamma(x)$ is strictly log-convex with its minimum lying between $1.46$ and $1.47$, implying that the mode of $f_k$ lies in $(0, 1.47)$, while when $B_k$ is negative, the mode of $f_k$ lies in $(0, e^{1-B_k/J})$, which is proved in Lemma \ref{lemma:mode-neg-B} in \S\ref{lemmas} of the Supplement.
For $x>0$, let 
\begin{align*}
    \Tilde{h}_k(x) & = \log \ \Tilde{f}_k(x) =  - J \,\log \,\Gamma(x) + (A-1) \, \log \, x - B_kx\,, \\
    \Tilde{h}^\prime_k(x) & =  - J \,\psi(x) + \frac{(A-1)}{x} - B_k
\end{align*}
denote the log density upto constants and its derivative with respect to $x$. 

For $N \in \mathbb{N}$, we consider $2N+2$ knot points denoted by $0 < m_{k,1}< \ldots < m_{k,N+1}< \ldots < m_{k,2N+2} < \infty$ which are selected in a way that they are concentrated around the mode of the density so as to ensure a tight upper envelope for our target density and hence a high acceptance probability.
More specifically, we choose the knots in the following manner :
\begin{enumerate}
    \item Set the central knot $m_{k,N+1}$ at the mode of $f_k$, which is obtained numerically by solving for $\Tilde{h}^\prime_k(m_{k,N+1}) = 0$.
    \item Set the last knot $m_{k, 2N+2} = (m_{k,N+1} + 1.5)\,\mathds{1}_{\{B_k>0\}} + e^{1-B_k/J} \, \mathds{1}_{\{B_k<0\}}$.
    \item Set the first and last but one knots as $m_{k,1} = m_{k,N+1}/2$ and $m_{k, 2N+1} = (m_{k,N+1}+ m_{k,2N+2})/2$.
    \item The remaining $N-1$ knots to the left and right of the mode are chosen as equidistant points between $m_{k,1}$ and $m_{k,N+1}$, and $m_{k,N+1}$ and $m_{k, 2N+1}$ respectively.
\end{enumerate}

Equation of the tangent line of $\Tilde{h}_k$ at the point $m_{k,i}$ is given by
\begin{equation*}
\label{tangent}
    \nu_{k,i}(x) = a_{k,i} + \lambda_{k,i} (x - m_{k,i}), \quad i = 1, \ldots, 2N+2
\end{equation*}
where $a_{k,i} = \Tilde{h}_k(m_{k,i})$ and $\lambda_{k,i} = \Tilde{h}^\prime_k(m_{k,i})$. 
Clearly, $\lambda_{k,i} > 0$ for $i \leq N$,
$\lambda_{k, N+1} = 0$ and $\lambda_{k,i} < 0$ for $i \geq N+2$.
Points of intersection of the tangent lines, $\nu_{k,i}$ and $\nu_{k,i+1}$ is given by
\begin{equation*}
\label{points}
    q_{k,i} = \frac{a_{k, i+1} - a_{k,i} + m_{k,i} \lambda_{k,i} - m_{k, i+1} \lambda_{k,i+1}}{\lambda_{k,i} - \lambda_{k,i+1}}, \quad \quad i = 1,\ldots, 2N+1.
\end{equation*}
Let $q_{k,0}=0$, $q_{k,2N+2} = \infty$ and $\nu_k(x) : = \sum_{i=1}^{2N+2} \nu_{k,i}(x) \ \mathds{1}_{\left\{x\in [q_{k,i-1}, \,q_{k,i})\right\}}$.
Since $\Tilde{h}_k$ is concave in nature and concave functions lie below their tangent lines, we have $\Tilde{f}_k(x) \leq e^{\,\nu_k(x)}$, which provides a piece-wise upper bound $\Tilde{g}_k$ for $\Tilde{f}_k$,
\begin{equation*}
\label{cover_B_k}
    \Tilde{f}_k(x) \ \leq \ \Tilde{g}_k(x) : = 
    \sum_{i=1}^{2N+2} \Tilde{g}_{k,i} (x)
\end{equation*}
where $\Tilde{g}_{k,i}(x) = e^{\nu_{k,i}(x)} =  e^{a_{k,i} + \lambda_{k,i} (x - m_{k,i})}\ \mathds{1}_{\left\{x\in [q_{k,i-1}, \,q_{k,i})\right\}}, \ i = 1, \ldots, 2N+2$.
The final cover density $g_k = \Tilde{g}_k/ C_{g_k}$ has the normalizing constant $C_{g_k} = \int_0^\infty \Tilde{g}_k(x)dx= \sum_{i = 1}^{2N+2}C_{g_{k,i}}$, where 
\begin{equation*}
    C_{g_{k,i}} = \int_{q_{k,i-1}}^{q_{k,i}} \Tilde{g}_{k,i}(x) \ dx = 
    \begin{cases}
        e^{a_{k, N+1}}\ (q_{k,N+1} - q_{k,N}), & i = N+1 \\
        {\lambda^{-1}_{k,i}}\ {e^{a_{k,i} - m_{k,i} \lambda_{k,i}}} \left( e^{q_{k,i} \lambda_{k,i}} - e^{q_{k,i-1} \lambda_{k,i}}\right), & i \neq N+1
    \end{cases}
\end{equation*}
The cover density $g_k$ can then be written as a mixture of $2N+2$ densities,  $g_{k,i} =  $,
\begin{equation*}
    g_k(x) = \sum_{i=1}^{2N+2}\frac{C_{g_{k,i}}}{C_{g_k}} \ g_{k,i}(x) ,
\end{equation*}
\begin{equation*}
    g_{k,i}(x) = \frac{\Tilde{g}_{k,i}(x)}{{C_{g_{k,i}}}} = 
    \begin{cases}
        (q_{k,N+1} - q_{k,N})^{-1}\ \mathds{1}_{\{x \in [q_{k, N}, \,q_{k, N+1})\}}, & i = N+1 \\
        ({e^{q_{k,i}\lambda_{k,i}} - e^{q_{k,i-1}\lambda_{k,i}}})^{-1}{\lambda_{k,i}}\ e^{\lambda_{k,i}x}\ \mathds{1}_{\{x \in [q_{k,i-1},\, q_{k,i})\}}, & i \neq N+1\\
    \end{cases}
\end{equation*}

\vspace{1ex}

\begin{figure}[ht]
    \centering
    \includegraphics[width=15cm]{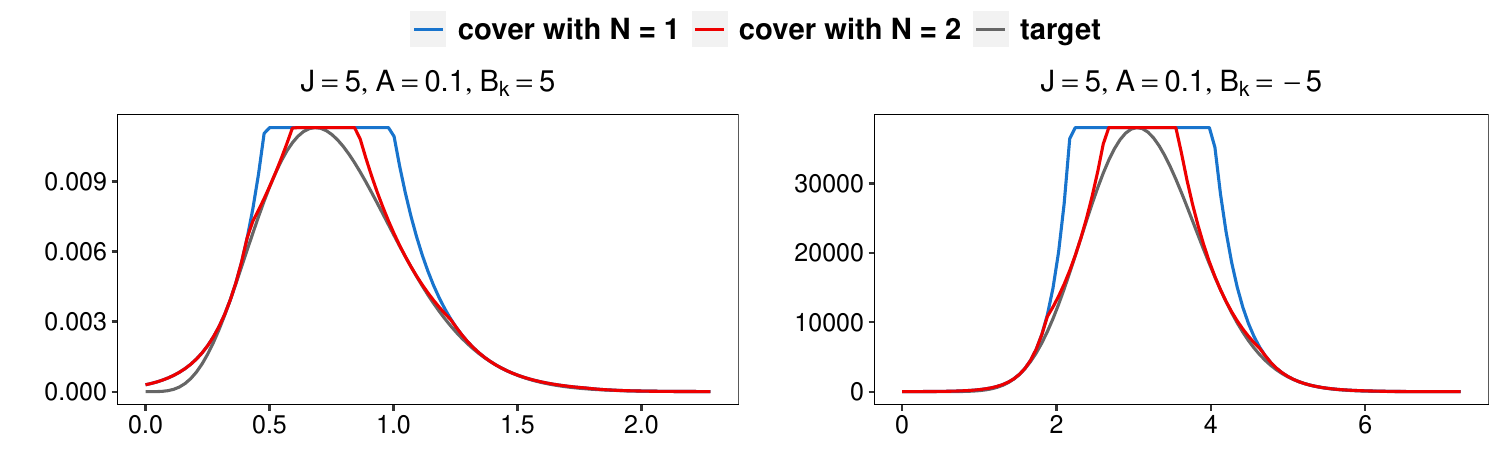}
    \caption{Plot of the unnormalized target $\Tilde{f}_k$ and the constructed unnormalized cover $\Tilde{g}_k$ with $N = 1$ (blue) and $2$ (red) for one choice of $J$, $A$, and positive (left) and negative (right) $B_k$.}
    \label{fig:knots}
\end{figure}

Figure \ref{fig:knots} shows the plot of $\Tilde{f}_k$ and $\Tilde{g}_k$ for one choice of parameters $J$, $A$ and a positive and negative $B_k$ where $N$ is chosen as $1$ and $2$ \ie $4$ and $6$ knot points are used for constructing the piecewise tangent lines, clearly demonstrating the high acceptance rate that the tight cover ensures. Plots for several choices of parameters by taking $N=1$ and empirical illustrations on the acceptance rate are provided in \S\ref{illustrations-AR} of the Supplement. Benefits of using such a rejection sampler are multi-fold. We get exact posterior samples of $\ubeta$ without the involvement of any tuning parameter. More importantly, due to the conditional independence of $t_k$, there is no significant computational hassle if the dimension of $\ubeta$ is high. Our devised blocked Gibbs sampler is presented in Algorithm \ref{alg:BGS}.  




\RestyleAlgo{ruled}
\begin{algorithm}[ht]
  \caption{The blocked Gibbs sampler cycles through the following steps.\\
  Here, $f$ and $h$ denote the densities corresponding to $F$ and $H$ respectively and $i = 1, 2, \ldots, n_j$, $j = 1, 2, \ldots, J$ and $k = 1, 2, \ldots L$, unless otherwise specified.
  \label{alg:BGS}}
  Update $\uz$: Sample $z_{ji} \sim \upi_{ji}^{*}$, where $\upi_{ji, k}^{*} = \pi_{jk}f(x_{ji} \mid \phi_k) \big/ \sum_{l=1}^L \pi_{jl}f(x_{ji} \mid \phi_l)$.\\
  Update $\uphi$:
  Let $\{z_1^{*}, z_2^{*}, \ldots, z_m^{*} \}$ denote the set of current unique values of $\uz$.\\
  ~~~~~~~ For $k \in \uz \setminus \{z_1^{*}, z_2^{*}, \ldots, z_m^{*} \}$, sample $\phi_k \sim H$.\\
  ~~~~~~~ For $k \in \{z_1^{*}, z_2^{*}, \ldots, z_m^{*} \}$, sample $\phi_k \sim h(\phi_k) \prod_{ji:z_{ji}=k} f(x_{ji} \mid \phi_k)$.\\
  Update $\upi$:
  Sample $\upi_j \sim \text{Dir}(\un^\prime_j +\ut)$, where $n^\prime_{jk} = \sum_{i=1}^{n_j} \mathds{1}_{\{z_{ji}=k\}}$.\\
  Update $\ut$ and $\ubeta$ :\\
  ~~~~~~~ Sample $t_k \sim f_k$ for each $k$ using the rejection sampler and set $\beta_k = t_k / \sum_l t_l$.\\
  Update $\uu$: Sample $u_j \sim \text{Gamma} \left(\sum_k t_k, 1 \right)$. 
\end{algorithm}

\subsection{Potential restrictiveness of the prior on the shared concentration parameter}

One key observation in our blocked Gibbs sampler is that we do not have flexibility in choosing the shape parameter $a_0$ of the gamma prior on the shared concentration $\alpha_0$, as our algorithm thrives by fixing it at the concentration $\gamma$ of the global DP, which may be of potential concern to practitioners. To respond to this, we first note that while introducing HDP, \cite{Teh2006HierarchicalDP} specify vague gamma priors on the concentration parameters, following a similar specification for the concentration of a DP in \cite{Escobar1995BayesianDE}. For our formulation, the rate parameter $b_0$ of the gamma prior can be chosen to lie in $(0, 1)$ which inflates the prior variance of $\alpha_0$, adhering to the recommendations of \cite{Teh2006HierarchicalDP}.  Additionally, sensitivity analyses carried out in \S\ref{sensitivity} of the Supplement illustrate the robust performance of our algorithm across various choices of $(\gamma, b_0)$.

\section{Simulation Study}
\label{simulation}
To investigate the performance of our blocked Gibbs sampler, referring to it as BGS henceforth, we consider a simulation setup that assesses both statistical accuracy in terms of clustering and density estimation, as well as algorithmic accuracy in terms of mixing behavior and computational efficiency. 
\cite{Wu_HDP} compared the performance of the exact slice sampler \citep{Amini2019ExactSS} with the CRF based collapsed sampler \citep{Teh2006HierarchicalDP} and the split-merge algorithm \citep{WangBlei2012}, where the slice sampler is shown to surpass the others. Code for the other algorithms discussed in \S\ref{overview_algorithms} are not publicly available.
Hence, we compare our algorithm with the CRF based collapsed sampler \citep{Teh2006HierarchicalDP}, the slice sampler \citep{Amini2019ExactSS} and the unmarginalized blocked Gibbs sampler outlined in \S3.1 of \cite{Amini2019ExactSS} where we use finite truncations of the global and local atoms instead of slicing techniques. The algorithms are referred to as CRF, SS and uBGS respectively. Note that while CRF and SS are exact samplers, BGS and uBGS are truncated and are hence considered approximate algorithms.

We specify $J = 3$ groups and consider a Gaussian mixture model having $4$ true components, the means of which are taken to be (i) $\uphi^{0} = (-3, -1, 1, 3)$ to allow for overlap between the adjacent densities, and (ii) $\uphi^{0} = (-6, -2, 2, 6)$ where the densities are well separated, with common precision $\tau = 1$ in both cases.  The mixture weights are chosen as $\upi^0_1 = (0.5, 0.5, 0, 0)$, $\upi^0_2 = (0.25, 0.25, 0.25, 0.25)$ and $\upi^0_3 = (0, 0.1, 0.6, 0.3)$ to accommodate different mixture distributions for each group. Considering equal sample sizes $n_j = n$ for each group, we generate the true cluster labels $z^0_{ji} \sim \upi_j$ and the observations $x_{ji} \sim \mathcal{N}(\phi^0_{z^0_{ji}}, \tau^{-1})$, $i = 1, 2, \ldots, n_j$, $j = 1, 2, \ldots, J$. 
The true density of group $j$ is given by
$$f^{\,0}_j(y) = \sum_{k=1}^4 \pi^{0}_{jk} \ \mathcal{N}(y\,;\, \phi_{k}^{0}, \tau^{-1}), \ y \in \mathbb{R},$$
where $\mathcal{N}(\cdot\,;\, \phi, \tau^{-1})$ denotes a normal density with mean $\phi$ and precision $\tau$. We assume a $\mathcal{N}(\xi, \lambda^{-1})$ conjugate prior on each $\phi_k$ and set $\xi = 0$ and $\lambda = 1$. 
The hyperparameters $\gamma$ and $b_0$ are taken as $1$ and $0.1$ respectively while the truncation level $L$ is fixed at $10$. Sensitivity analyses conducted in \S\ref{sensitivity} of the Supplement demonstrate robustness of our algorithm across varying choices of $(\gamma, b_0)$ and $L$. Finally, we choose $N = 1$ \ie our rejection sampler considers $4$ knot points for constructing the piecewise tangent lines.

For CRF, we noted robust performance across different choices of initial number of clusters and consequently set it at $L$ for our simulations. For uBGS, truncation level for both the global and local clusters are set at $L$. Hyperparameter specifications are kept the same across all competing algorithms.
Since SS does not assume a prior on $\alpha_0$, we explore its performance by considering three different choices for it as 0.1, 1 and mean of our chosen gamma prior \ie $10$.
We run all the algorithms considering $n = 50, 100, 200$ and collect $M = 1000$ posterior samples after a burn-in of $2000$ samples for each $n$. Posterior estimates are summarized over $50$ simulation replicates. Code for implementing our method is available at the GitHub repository, \href{https://github.com/das-snigdha/blockedHDP}{das-snigdha/blockedHDP}. 

\subsection{Statistical Accuracy}
Statistical accuracy of BGS is measured by comparing the performance in recovering the true cluster labels and estimating the density for each group, with that of the competing algorithms, using the adjusted Rand index (ARI) \citep{Rand1971ObjectiveCF, Hubert1985ComparingP} and mean integrated squared error (MISE) as the respective discrepancy measures. To circumvent the issue of label switching arising in posterior samples of parameters in mixture models, the posterior estimate of cluster labels is obtained by employing the least-squares clustering method \citep{dahl_2006}.
Let $\hat{\uz}_{j}$ denote the estimated cluster labels for group $j$ and $\hat{\uz} = (\hat{\uz}_1, \ldots, \hat{\uz}_J )$ denote the estimated global cluster labels. We evaluate $\text{ARI} (\hat{\uz}_j, {\uz}^{0}_j )$ and $\text{ARI} (\hat{\uz} , {\uz}^{0} )$ to assess the local and global clustering performances.
For density estimation, we consider $100$ equidistant grid points $\{y_h : h = 1, 2, \ldots, 100 \}$ in $[x_{\min} - 1, x_{\max} + 1]$, where $x_{\min} = \min \{x_{ji} : i, j\}$, $x_{\max} = \max \{x_{ji} : i, j\}$, and get the posterior density estimate at each grid point $y_h$ for group $j$ as $$\hat{f}_j(y_h) = \frac{1}{M}\sum_{m = 1}^{M}\sum_{k=1}^L \pi^{(m)}_{jk} \ \mathcal{N} (y_h\,;\, \phi_{k}^{(m)}, \tau^{-1}),$$
where $\uphi^{(m)}$ and $\upi^{(m)}$ denote the $m^{th}$ posterior samples of $\uphi$ and $\upi$ respectively.
The MISE of $\hat{f}_j$ is calculated as $\text{MISE}\, \big(\hat{f}_j \big) = ({1}/{100}) \sum_{h=1}^{100} \left\{ \hat{f}_j (y_h) - f_j^{\,0} (y_h)\right\}^2$.

\begin{figure}[htp]
     \centering
     \begin{subfigure}[b]{0.93\textwidth}
         \centering
         \includegraphics[width=\textwidth]{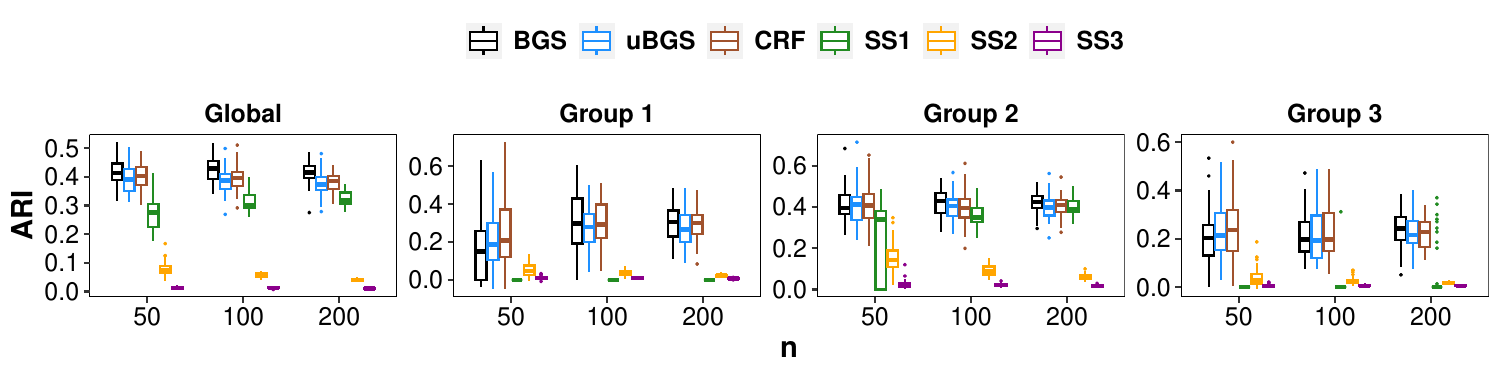}
         \caption{}
     \end{subfigure}
     \begin{subfigure}[b]{0.93\textwidth}
         \centering
         \includegraphics[width=\textwidth]{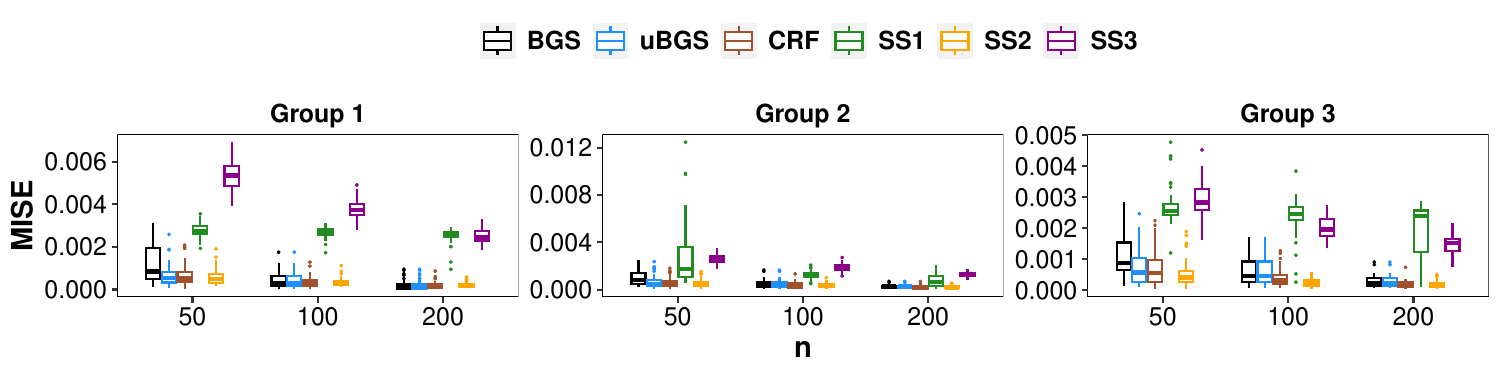}
         \caption{}
     \end{subfigure}
     \begin{subfigure}[b]{0.93\textwidth}
         \centering
         \includegraphics[width=\textwidth]{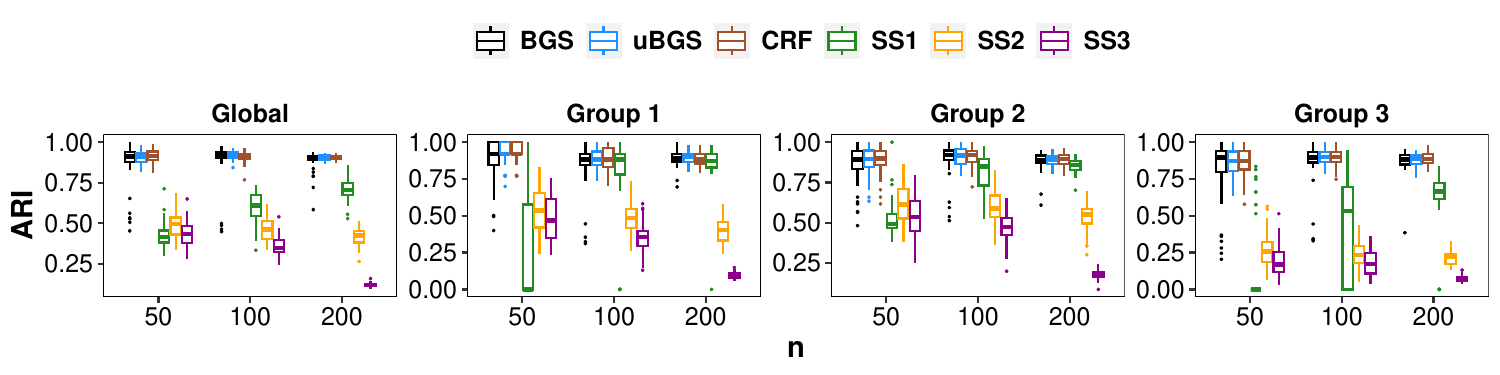}
         \caption{}
     \end{subfigure}
     \begin{subfigure}[b]{0.93\textwidth}
         \centering
         \includegraphics[width=\textwidth]{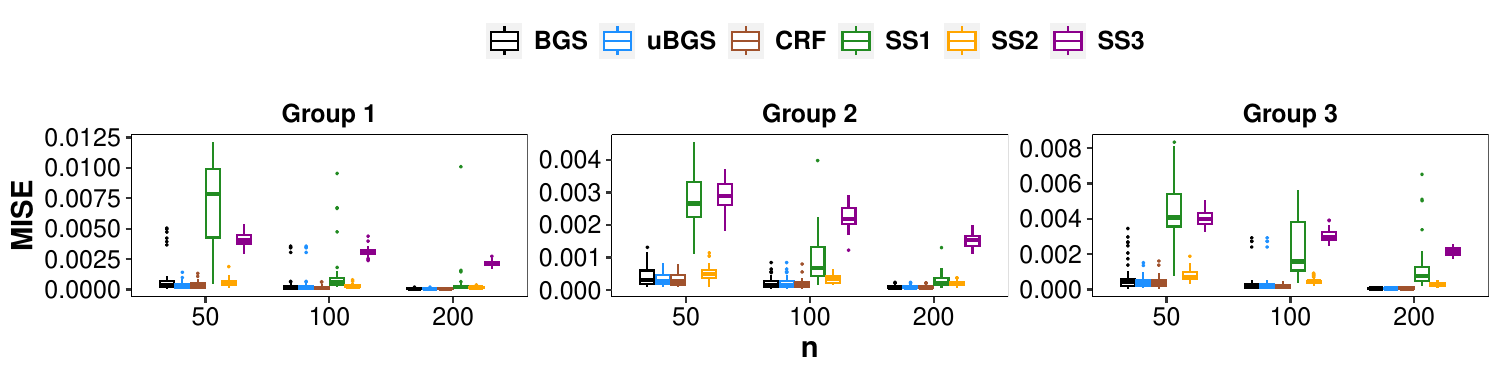}
         \caption{}
     \end{subfigure}
    \caption{Adjusted Rand indices -- (a),(c) of the estimated cluster labels and mean integrated squared error -- (b),(d) of the estimated densities, when true means of the Gaussian mixture are $\uphi^{0} = (-3,-1,1,3)$ --(a),(b) and $\uphi^{0} = (-6,-2,2,6)$ -- (c),(d). SS1, SS2, SS3 refer to the slice samplers with $\alpha_0$ chosen as 0.1, 1, 10 respectively. Boxplots show variation across 50 simulation replicates.}
    \label{fig:StatAccuracy}
\end{figure}

Figure \ref{fig:StatAccuracy} shows boxplots of the ARIs and the MISEs across 50 replicates. The ARIs remain generally low under the overlapping design while being closer to 1 under the well-separated design.
An overall decreasing trend in the MISE with increase in the sample size shows that posterior estimates of our algorithm are consistent. The clearly noticeable proximity in the ARI of cluster labels and the MISE of the density estimates obtained from BGS, uBGS and CRF suggests that these samplers provide very similar estimation performances. SS shows an overall worse performance, except for its density estimates consistently aligning with BGS when $\alpha_0$ is set to $1$. Additional plots showing the estimated cluster labels and densities in one of the simulation replicates are provided in \S\ref{addplots_estimation} of the Supplement.

\subsection{Algorithmic Accuracy}
\begin{figure}[htp]
     \centering
     \begin{subfigure}[b]{0.93\textwidth}
         \centering
         \includegraphics[width=\textwidth]{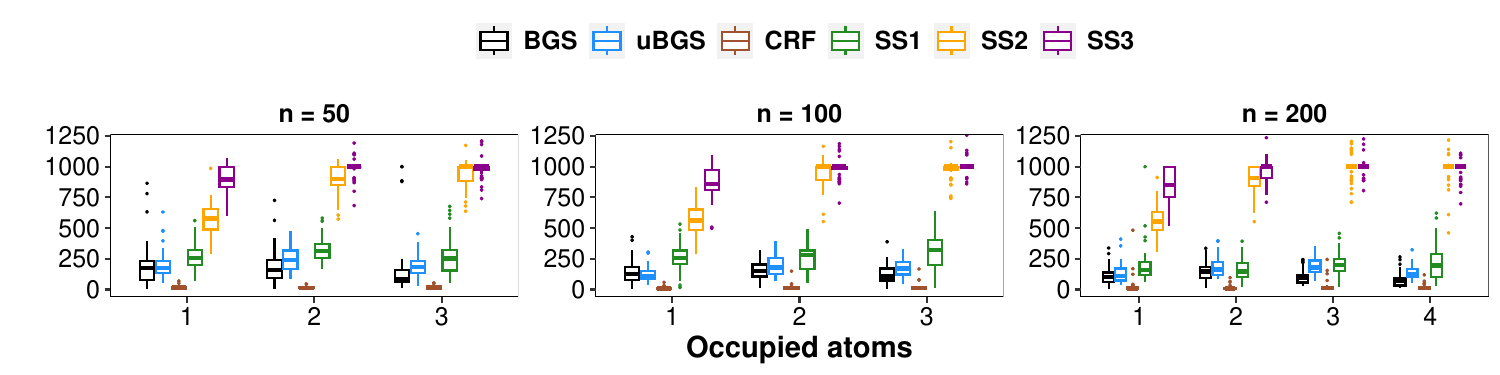}
         \caption{}
     \end{subfigure}
     \begin{subfigure}[b]{0.93\textwidth}
         \centering
         \includegraphics[width=\textwidth]{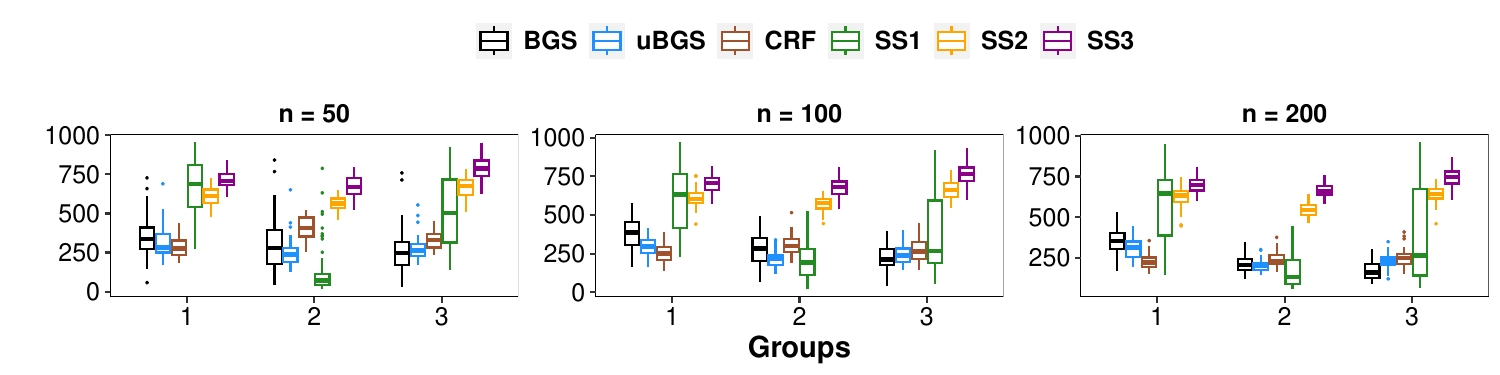}
         \caption{}
     \end{subfigure}
     \begin{subfigure}[b]{0.93\textwidth}
         \centering
         \includegraphics[width=\textwidth]{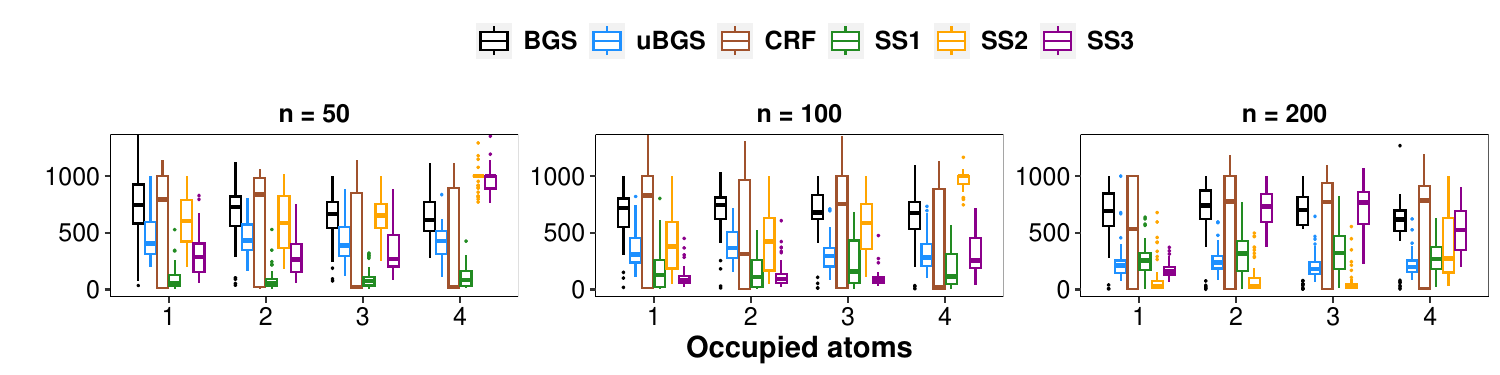}
         \caption{}
     \end{subfigure}
     \begin{subfigure}[b]{0.93\textwidth}
         \centering
         \includegraphics[width=\textwidth]{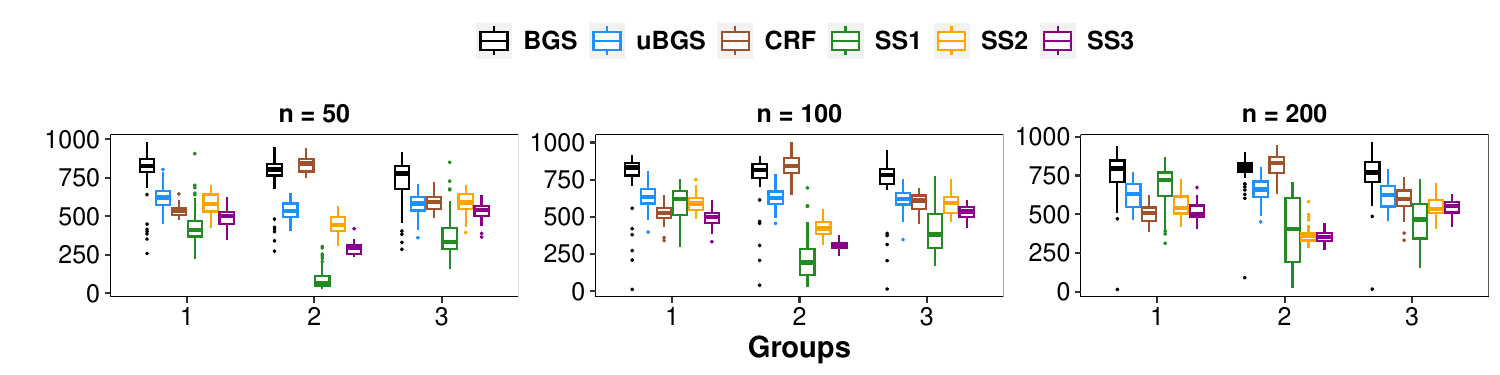}
         \caption{}
     \end{subfigure}
    \caption{Effective sample sizes of occupied atoms -- (a),(c) and estimated densities -- (b),(d), when true means of the Gaussian mixture are $\uphi^{0} = (-3,-1,1,3)$ -- (a),(b) and $\uphi^{0} = (-6,-2,2,6)$ -- (c), (d). SS1, SS2, SS3 refer to the slice samplers with $\alpha_0$ chosen as 0.1, 1, 10 respectively. Boxplots show variation across 50 simulation replicates.}
    \label{fig:Mixing}
\end{figure}

In the following, we discuss mixing behavior of the sampled atoms and the estimated densities, and the computational efficiency of our algorithm. 
While CRF allows dynamic estimation of the number of clusters and samples atoms for each occupied cluster, BGS, uBGS and SS on the contrary retain the unassigned clusters, the atoms corresponding to which are drawn independently from their prior. To bring all samplers on the same page, we define $\hat{L} = \min \{\hat{L}^{(m)}: m = 1, 2, \ldots, M\}$, where $\hat{L}^{(m)}$ denotes the number of estimated clusters by CRF in each posterior sample $m$,  and retain the first $\hat{L}$ atoms from the posterior samples of CRF. We then extract $\hat{L}$ atoms from BGS, uBGS and SS in decreasing order of their corresponding cluster occupancy to ensure that the atoms drawn from their prior are dismissed. 
Figure \ref{fig:Mixing} shows boxplots of the effective sample sizes (ESS) of the occupied atoms and estimated densities, for each $n$ across 50 replicates. The plotted ESS of the density estimates is an average of element-wise ESS for the 100 grid points.  Note that ESS can be higher than the number of MCMC samples in the presence of negative autocorrelation.


Under the overlapping design, BGS, uBGS and CRF have comparable ESS for the density estimates of each group. BGS and uBGS exhibit comparable ESS for each of the atoms corresponding to the occupied clusters, that are higher than those of CRF, for each $n$. The better mixing behavior of BGS and uBGS can be attributed to their ability to update parameters in blocks as opposed to one-coordinate-at-a-time updates for parameters in CRF, and is consistent with the performance of the blocked Gibbs sampler for a DP \citep{Ishwaran2001GibbsSM}. Under the well-separated design, BGS has uniformly higher ESS for density estimates as well as occupied atoms, compared to uBGS. Sampling the local cluster labels in addition to the global labels in uBGS incurs higher autocorrelation and results in a slower mixing, as opposed to updating only the global labels in BGS. 
CRF is seen to have high variability in the ESS of occupied atoms across replicates under the well separated design, with the median being high in most of the cases. When the clusters are well separated, CRF often captures the true clusters quickly and samples the atoms independently for those clusters, consequently  exhibiting a high ESS in such replicates, while having a low ESS in replicates when it fails to do so. 

Under the overlapping design, all three implementations of SS have considerably higher ESS for the occupied atoms. SS also has higher ESS for the estimated densities, except that for group 2 when $\alpha_0$ is set at $0.1$. This behavior is however not consistent with what is observed under well separated true clusters. 
 It is important to note that ESS gives us the number of independent samples that have the same estimation efficiency as the given auto-correlated MCMC samples. SS shows a tendency to over-smooth the local modes of the estimated densities \ie it fails to capture all the distinct local modes (as evident in the plots of \S\ref{addplots_estimation}) and hence exhibits a higher ESS at the cost of inadequate estimation performance, compared to the other samplers. Evidently, there is a drop in its ESS under the well-separated design where both clustering and density estimation is less challenging, than under the overlapping design.
 

 \begin{figure}[htp]
    \centering
    \begin{subfigure}[b]{0.33\textwidth}
         \centering
         \includegraphics[width=\textwidth]{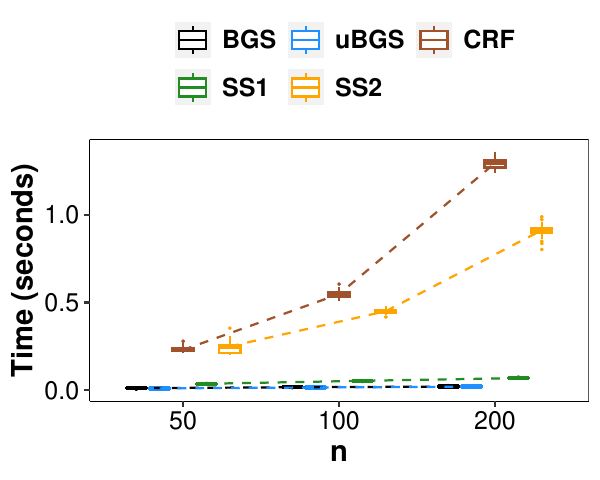}
         \caption{}
     \end{subfigure}
     \hspace{1.5cm}
     \begin{subfigure}[b]{0.33\textwidth}
         \centering
         \includegraphics[width=\textwidth]{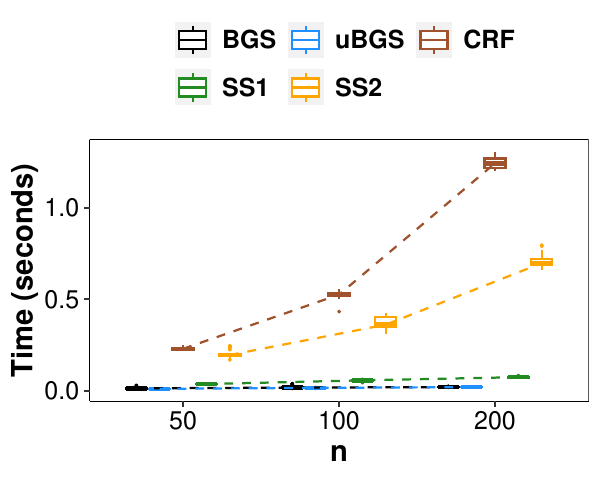}
         \caption{}
     \end{subfigure}
    \caption{Average computation time (in seconds) per MCMC iteration, when true means of the Gaussian mixture are (a) $\uphi^{0} = (-3,-1,1,3)$ and (b) $\uphi^{0} = (-6,-2,2,6)$. SS1 and SS2 refer to the slice samplers with $\alpha_0$ chosen as 0.1 and 1, respectively. Boxplots show variation across 50 simulation replicates, along with trajectories of the median time.}
    \label{fig:Time}
\end{figure}

A remarkable gain in computational efficiency in using the blocked samplers as compared to CRF, is clearly evident in Figure \ref{fig:Time} that shows boxplots of the average computation time per MCMC iteration for the samplers across 50 replicates with increasing $n$. 

Since the cluster assignments for the blocked samplers in the truncated model factorize over the observations in each group, the likelihood evaluation $f(x_{ji}\,|\, \phi_k)$ for each of the $L$ atoms incurs a computational cost of $\mathcal{O}\left(n_0L \right)$, where $n_0 = \sum_{j=1}^J n_j$ denotes the total sample size. 
Cluster assignments of a data point under the CRF scheme is conditional on all other data points. Consequently, the likelihood evaluation for each observation in each group incurs an additional $(n_0-1)$ computations, resulting in an overall cost of $\mathcal{O}\left(n_0^2L^\prime \right)$, where $L^\prime$ is the maximum number of clusters assigned by the algorithm. 

The computation time for SS is seen to exhibit an increase when a higher value is assigned to $\alpha_0$. Consequently, the computation time for SS with $\alpha_0$ set at 10 is deferred to \S\ref{addplots_time} of the Supplement, to retain visual clarity in the comparison of all algorithms.
Trajectories for the median of the average computation time demonstrate the scalability of BGS (comparable to that of uBGS) with increasing $n$, as compared to CRF. The computation time can be further accelerated by trivial parallelization. Moreover, the posterior updates factorize over the $J$ groups, making the sampler suitable for applications with large sample sizes as well as large number of groups.

\vspace{0.3cm}

\begin{table}[ht]
    \centering
    \begin{tabular}{l c c c c c c c c} 
    \hline
    & \multicolumn{4}{c}{Overlapping design} & \multicolumn{4}{c}{Well-separated design}\\
    & BGS & uBGS & CRF & SS & BGS & uBGS & CRF & SS \\ 
    \hline
    Clustering (ARI) & $\checkmark $ & $\checkmark $ & $\checkmark$ &$\bm \cdot$ & $\checkmark$ & $\checkmark $ & $\checkmark$ & $\bm \cdot$  \\ 
    Density estimation (MISE) & $\checkmark$ & $\checkmark $ & $\checkmark$ & $\ast$ & $\checkmark$ & $\checkmark $ & $\checkmark$ & $\ast$ \\
    Mixing (ESS) & $\checkmark$ & $\checkmark $ & $\bm \cdot$ & $\bm \cdot$ & $\checkmark$ & $\bm \cdot $ & $\ast$ &$\bm \cdot$ \\
    Scalability (Computation time) & $\checkmark$ & $\checkmark $ & $\bm \cdot$ & $\ast$ & $\checkmark$ & $\checkmark $ &$\bm \cdot$ & $\ast$ \\ 
     \hline 
    \end{tabular}
    \caption{Summary of the statistical and algorithmic performances of the three algorithms, as seen in our simulations. The symbols indicate the following performances: $\checkmark$ favorable, $\bm \cdot$ unfavorable or uninterpretable, $\ast$ favorable in some implementations.}
    \label{tab:performance}
\end{table}




\bibliographystyle{apalike}
\bibliography{ref}

\newpage

\section*{\LARGE Supplementary Materials} 
This supplement contains an enlisting of blocked parameter updates of the truncated HDP mixture model, sampling scheme for the tilted gamma random variables, proof of auxiliary results, sensitivity analyses with respect to choices of hyperparameters and the truncation level, and additional plots depicting the computation time and estimation performance of the different posterior sampling algorithms. Code for implementing our method is available at the GitHub repository, \href{https://github.com/das-snigdha/blockedHDP}{das-snigdha/blockedHDP}.

\appendix
\section{Blocked parameter updates of the truncated HDP mixture model}
\label{full_cond}
Let $F(\theta)$ have density $f(\cdot\mid\theta)$ and $H$ have density $h(\cdot)$. Likelihood and prior specifications from model \eqref{finite_HDP}, along with a Gamma $(a_0, b_0)$ prior on $\alpha_0$, are as follows,
\begin{align*} 
    \ubeta \mid \gamma & \sim \text{Dir}\left({\gamma}/{L}, \ldots, {\gamma}/{L}\right)
    & \phi_k \mid H &\sim H  \\
    \upi_j \mid \beta, \alpha_0 & \sim \text{Dir}\left(\alpha_0 \ubeta \right)       
    &  z_{ji} \mid \upi_j &\sim \upi_j\\
    \alpha_0 \mid a_0, b_0 &\sim \text{Gamma }(a_0, b_0) & x_{ji} \mid z_{ji}, \{\phi_k \}_{k=1}^L &\sim F(\phi_{z_{ji}}) 
\end{align*}
The full conditional distributions for the parameters are given by 

\begin{enumerate}
    \item \textbf{Sampling} $\uphi$. For each $k=1, 2,\ldots,L$,
    \begin{align*}
    p(\phi_k\mid \ux, \uz, \upi, \ubeta, \alpha_0) & \propto  \ h(\phi_k) \prod_{ji:  z_{ji}=k}  f(x_{ji}\mid \phi_k)
    \end{align*}
    If $H$ is chosen conjugate to $F$, then we have independent conjugate updates for $\phi_1, \phi_2, \ldots, \phi_L$.
    
    \item \textbf{Sampling} $\uz$. For each $i = 1, 2, \ldots, n_j$ and $j = 1, 2, \ldots, J$,
    \begin{align*}
        z_{ji} \mid \ux, \uphi, \upi, \ubeta, \alpha_0 \sim \upi^{*}_{ji}
    \end{align*}
    where $\pi^{*}_{ji,k} = {\pi_{jk}f(x_{ji}\mid \phi_k)}\big/{\sum_{l=1}^L \pi_{jl}f(x_{ji}\mid \phi_l)}$ for each $k$.
    
    \item \textbf{Sampling} $\upi$. For each $j = 1, 2, \ldots, J$, 
    \begin{equation*}
        \upi_j \mid \ux, \uz, \uphi, \ubeta, \alpha_0 \ \sim \ \text{Dir}(\un^\prime_j + \alpha_0\ubeta)
    \end{equation*}
    where $\un^\prime_j = (n^\prime_{j1}, n^\prime_{j2}, \ldots, n^\prime_{jL}), \ n^\prime_{jk} = \sum_{i=1}^{n_j} \mathds{1}_{\{z_{ji}=k\}}$ for each $k$.
    
    \item \textbf{Sampling} $\ubeta$. 
    \begin{equation*}
        p(\ubeta \mid \ux, \uphi, \uz, \upi,\alpha_0 ) \propto \ \prod_{k=1}^L \ \frac{1}{ \Gamma(\alpha_0 \beta_k)^J} \ \big(\prod_{j=1}^J \pi_{jk}\big)^{\alpha_0 \beta_k}\, \beta_k^{\,\frac{\gamma}{L} -1}\ \mathds{1}_{\left\{ \ubeta \,\in \, \mathcal{S}^{L-1} \right\}}
    \end{equation*}
    where $\mathcal{S}^{L-1}$ denotes the $L$-dimensional simplex.

    \item \textbf{Sampling} $\alpha_0$. 
    \begin{equation*}
        p(\alpha_0 \mid \ux, \uphi, \uz, \upi,\ubeta)\propto \alpha_0 ^{a_0-1} e^{-b_0\alpha_0} \ \frac{\Gamma(\alpha_0)^J}{\prod_{k=1}^L\Gamma(\alpha_0 \beta_k)^J}\ \bigg[ \prod_{k=1}^L \big(\prod_{j=1}^J \pi_{jk}\big)^{\beta_k} \bigg]^{\alpha_0} \ \mathds{1}_{\{ \alpha_0 > 0\}}
    \end{equation*}
\end{enumerate}

\vspace{2ex}

\section{Rejection sampler for the tilted gamma random variables}
\label{samplerforbeta}

In the following, we describe the sampling scheme to get exact samples from the titled gamma density $f_k$, $k=1,2,\ldots, L$, given by 
\begin{equation*}
\begin{split}
    f_k(x) & \propto \frac{1}{ \Gamma(x)^J} \ x ^{\, A -1} \ e^{-B_k x}   \ \mathds{1}_{\left\{ x > 0 \right\}} 
\end{split}
\end{equation*}
where $\ A = \gamma/L $, $ \ B_k = b_0 - \sum_{j=1}^J\text{log } \pi_{jk} - \sum_{j=1}^J \text{log }u_j$. Let $\Tilde{f}_k(x) = \frac{1}{ \Gamma(x)^J} \ x ^{\, A -1} \ e^{-B_k x} \ \mathds{1}_{\left\{ x > 0 \right\}}$ and $C_{f_k} = \int_0^\infty \Tilde{f}_k(x) \, dx$ denote the density upto constants and its normalizing constant, respectively.

\subsection{Rejection sampling algorithm}
\label{rejectionsampler}
We briefly mention the steps of a rejection sampler for getting exact samples from $f_k$. Let $g_k = \Tilde{g}_k/ C_{g_k}$ be a density on $\mathbb{R}^{+}$ that can be sampled from and $\Tilde{f}_k(x) \leq \Tilde{g}_k(x)$ for all $x$ in the domain of $f_k$, then
a rejection sampling algorithm proceeds as follows:
\begin{enumerate}
    \item 
    Draw $s \sim g_k$ and $u \sim \text{Uniform }(0,1)$ independently.
    \item 
     Accept $s$ as a sample from $f_k$ if $u \ \leq \ \Tilde{f}_k(s)/\Tilde{g}_k(s)$. Otherwise, return to Step 1.
\end{enumerate}

\begin{figure}[htp]
     \centering
     \begin{subfigure}[b]{0.95\textwidth}
         \centering
         \includegraphics[width=\textwidth]{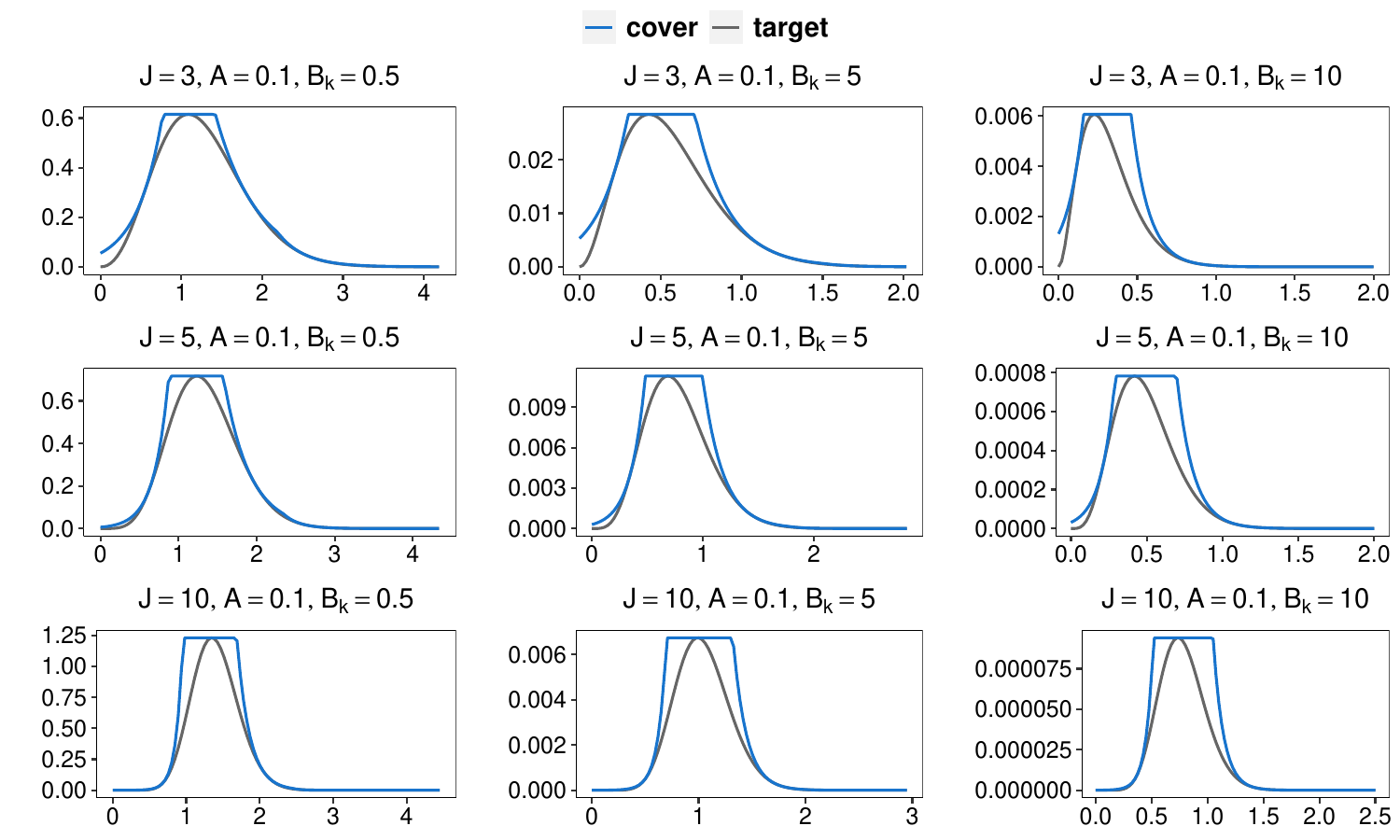}
         \caption{}
     \end{subfigure}
     \begin{subfigure}[b]{0.95\textwidth}
         \centering
         \includegraphics[width=\textwidth]{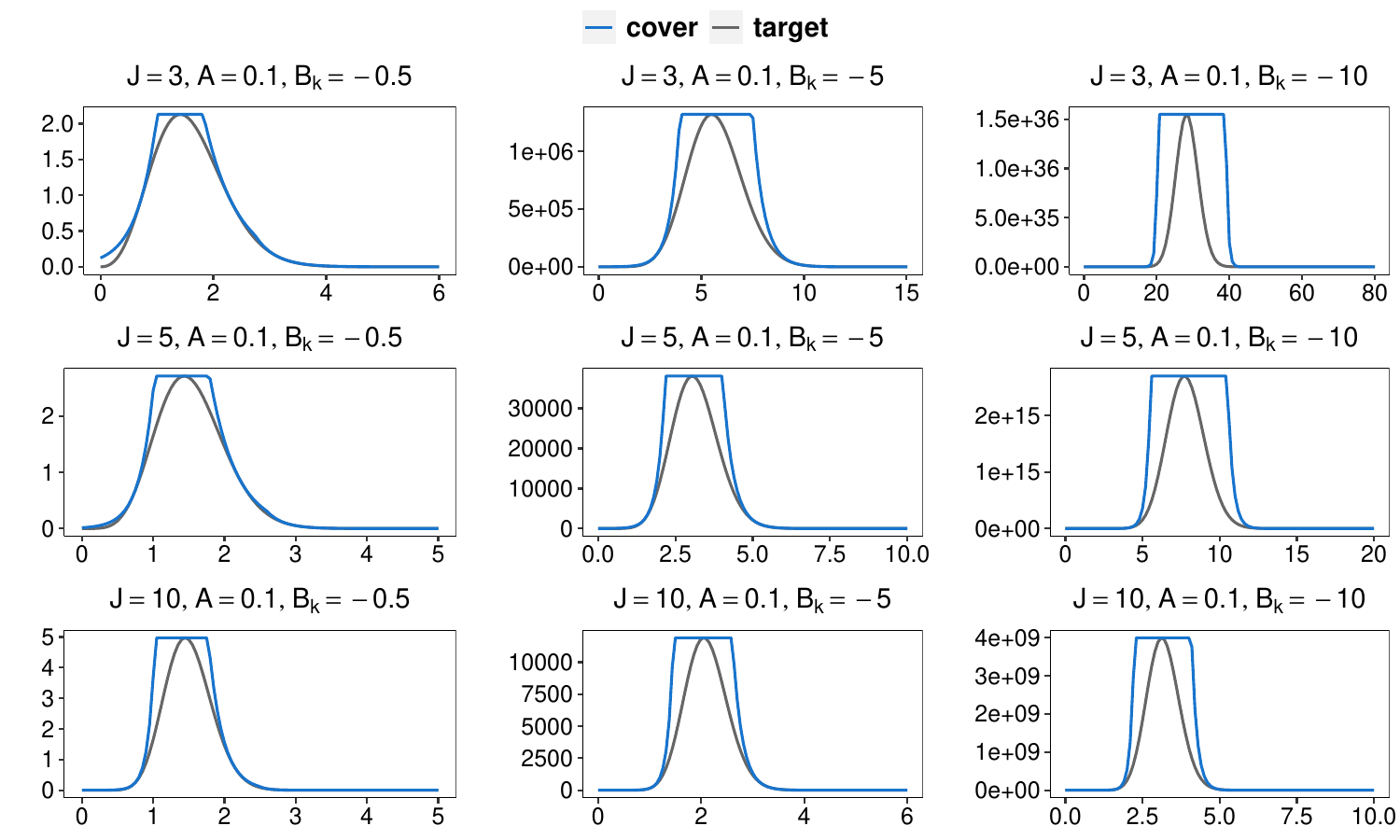}
         \caption{}
     \end{subfigure}
    \caption{Plot of the target $\Tilde{f}_k$ and the constructed cover $\Tilde{g}_k$ with $N = 1$ for several choices of $J$, $A$ and $B_k$ when (a) $B_k > 0$, (b) $B_k<0$.}
    \label{fig:covers}
\end{figure}

\subsection{Sampling from the cover density}

In this section, we describe the procedure to draw samples from the mixture density, $g_k$. For this purpose, let $w_{k,i} = C_{g_{k,i}}/C_{g_k}$, $i = 1, \ldots, 2N+2$ denote the mixture weights. Also, let $w_{k,0} = 0$. A sample $s$ from $g_k$ is obtained as follows:
\begin{enumerate}
    \item Generate a random variable $u \sim \text{Uniform }(0,1)$.
    \item If $u \in \left[\sum_{j=0}^{i-1} w_{k,j}, \sum_{j=0}^i w_{k,j} \right)$, then draw $s \sim g_{k,i}$, $i = 1, \ldots, 2N+2$.
\end{enumerate}

We then proceed to describe the inverse cdf method to get samples from $g_{k,i}$, $i=1, \ldots, 2N+2$.
First, we note that since $m_{k, N+1}$ is the mode of $f_k$, $\lambda_{k,N+1} = 0$ and $g_{k,N+1}$ is  a $U(q_{k,N}, q_{k,N+1})$ density which can be directly sampled from.
To get a sample $s$ from $g_{k,i}$, $i \neq N+1$,
which are exponential densities, the inverse cdf sampler proceeds as follows: 
\begin{enumerate}
    \item Draw $u \sim \text{Uniform }(0,1)$.
    \item Set $s = G_{k,i}^{-1}(u)$, where the cdf of $g_{k,i}$ and its corresponding inverse are given by
\begin{align*}
    G_{k,i}(x) &= \frac{e^{\lambda_{k,i} x} - e^{\lambda_{k,i} q_{k,i-1}}}{e^{\lambda_{k,i} q_{k,i}} - e^{\lambda_{k,i} q_{k,i-1}}} \mathds{1}_{\{x \in [q_{k,i-1},\, q_{k,i})\}} \\[5pt] 
    G_{k,i}^{-1}(u) &= \lambda_{k,i}^{-1}\ \log \big\{u\, e^{\lambda_{k,i} q_{k,i}} + (1-u)\, e^{\lambda_{k,i} q_{k,i-1}}\big\}
\end{align*}
\end{enumerate}

\subsection{Empirical Illustrations}
\label{illustrations-AR}

\begin{figure}[ht]
     \centering
     \begin{subfigure}[b]{0.45\textwidth}
         \centering
         \includegraphics[width=\textwidth]{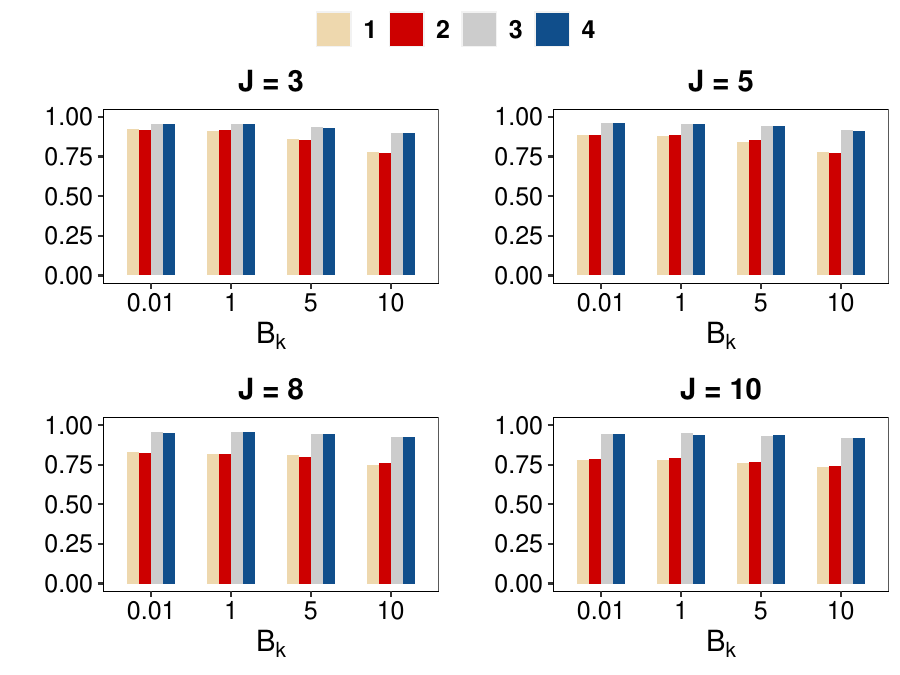}
         \caption{}
     \end{subfigure}
     \hspace{1ex}
     \begin{subfigure}[b]{0.45\textwidth}
         \centering
         \includegraphics[width=\textwidth]{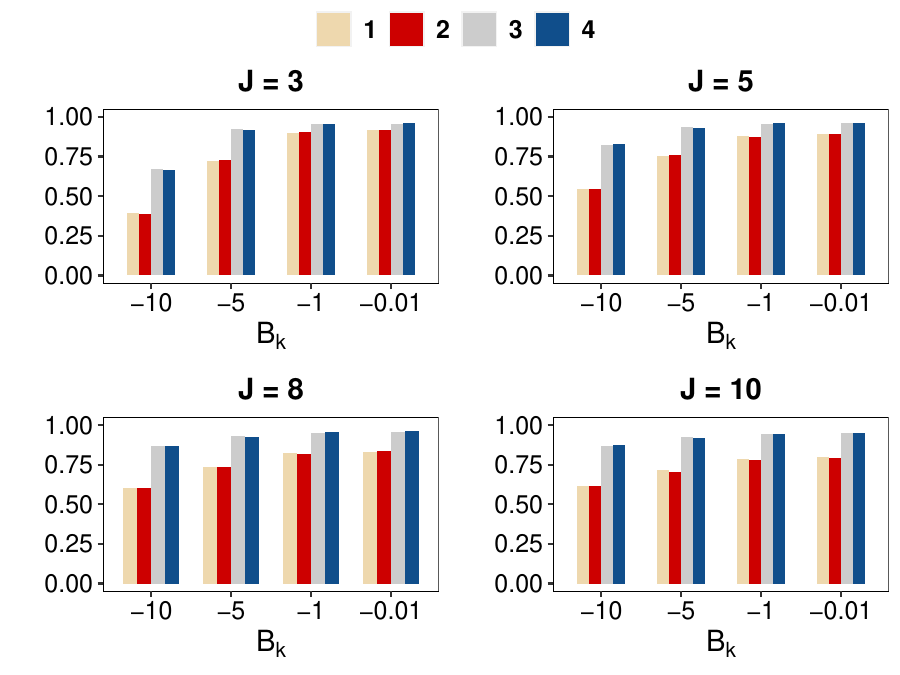}
         \caption{}
     \end{subfigure}
     \vspace{-0.1in}
    \caption{Barplots of acceptance probabilities of the rejection sampler with $N = 1, 2$, over various choices of $J$, $A$ and $B_k$ when (a) $B_k > 0$, (b) $B_k<0$. $1$ and $2$ in the legend correspond to $A = 0.01$ and $0.1$ respectively with $N = 1$, while $3$ and $4$ correspond to the same with $N = 2$. }
    \label{fig:AccRate}
\end{figure}

We investigate the performance of our proposed rejection sampler in terms of its acceptance probability. Figure \ref{fig:AccRate} shows barplots of the acceptance probabilities over varying choices of parameters $J$, $A$ and $B_k$ with $N$ chosen as $1$ and $2$. When $B_k$ is positive, an acceptance rate of at least $75\%$ is observed for all choices of $J$ and $A$ with $N$ fixed at $1$. A negative $B_k$ renders a slightly lower acceptance rate as compared to when it is positive. This is attributed to the right shifting of the mode of density for a negative $B_k$. Even so, we are able to get an acceptance rate of at least $40\%$ for all choices of $J$ and $A$ with $N$ fixed at $1$. It is important to note that $B_k$ involves a large positive the term, $-\log \big( \prod_{j=1}^J \pi_{jk} \big)$\,, $\pi_{jk}$ being the group-specific probabilities of each cluster, and hence the negative values it takes, remain small in magnitude, as seen under empirical runs of the sampler. The acceptance probabilities can be further boosted by choosing a higher value of $N$. Figure $\ref{fig:AccRate}$ demonstrates very high acceptance probabilities, obtained just by increasing $N$ from $1$ to $2$.

\section{Proof of auxiliary results}
\label{lemmas}

\begin{lemma}
\label{lemma:normalization}
A normalized vector $\uX$ of jointly distributed random variables, $\uY = \left(Y_1, Y_2, \ldots, Y_L \right)$ with density $f_{\uY}$ on $(0, \infty)^L$, $L \in \mathbb{N}$, has joint density of the form,
\begin{equation*}
    f_{X_1, X_2, \ldots, X_{L-1}}(x_1, x_2, \ldots, x_{L-1}) = \int_0^\infty y^{L-1}\, f_{\uY}(x_1y, x_2y, \ldots, x_Ly)\ dy
\end{equation*}
with $x_L = 1 - \sum_{i=1}^{L-1} x_i$.
\end{lemma}

\begin{proof}
\cite{Kruijer2010AdaptiveBD} derived the density for the vector of normalized $\uY$ for the case when $Y_1, Y_2, \ldots, Y_L$ are independent random variables, with $Y_i$ having density $p_i$ on $(0,\infty)$, $i = 1, 2, \ldots, L$. We provide a straightforward extension of their proof to the case when $Y_1, Y_2, \ldots Y_L$ are dependent random variables.

We begin with writing $\uX = (X_1, X_2, \ldots, X_L) =\left({Y_1}/{Y}, {Y_2}/{Y}, \ldots, {Y_L}/{Y} \right)$ with $Y = \sum_{i=1}^L Y_i$.
Then, for $(x_1, x_2, \ldots, x_L) \in \mathcal{S}^{L-1}$, we have
\begin{equation*}
\begin{aligned}
    & P \left( X_1 \leq x_1,\ X_2 \leq x_2, \ldots,\ X_{L-1}\leq x_{L-1} \right) \\
    & =  P \left( \frac{Y_1}{Y} \leq x_1,\ \frac{Y_2}{Y} \leq x_2, \ldots,\ \frac{Y_{L-1}}{Y}\leq x_{L-1} \right)\\
    & = \int_0^\infty \int_0^{x_1y} \int_0^{x_2y} \ldots \int_0^{x_{L-1}y} f_{\uY} \left(y_1, y_2, \ldots, y_{L-1}, y- \sum_{i=1}^{L-1}y_i \right) \ dy_{L-1} \, dy_{L-2}\ldots dy_1\, dy
\end{aligned}
\end{equation*}

The joint density is then given by
\begin{equation*}
    f_{X_1, X_2, \ldots, X_{L-1}}(x_1, x_2, \ldots, x_{L-1}) = \frac{\partial^{L-1}}{\partial x_1 \partial x_2 \ldots \partial x_{L-1} } \ P \left( X_1 \leq x_1,\ X_2 \leq x_2, \ldots,\ X_{L-1}\leq x_{L-1} \right)
\end{equation*}

Interchanging the $(L-1)$ derivatives and integrals,
we have 
\begin{align*}
    & f_{X_1, X_2, \ldots, X_{L-1}}(x_1, x_2, \ldots, x_{L-1}) \\
    & = \int_0^\infty \frac{\partial}{\partial x_1} \int_0^{x_1y}  \ldots \ \frac{\partial}{\partial x_{L-1}} \int_0^{x_{L-1}y} f_{\uY} \left(y_1, y_2, \ldots, y_{L-1}, y- \sum_{i=1}^{L-1}y_i \right) \ dy_{L-1} \ldots \, dy_1\, dy
\end{align*}

The proof follows by applying Leibniz integral rule for differentiation successively for 
$x_{L-1}$, $x_{L-2}, \ldots, x_1$. 

\end{proof}

\begin{lemma}
\label{lemma:log-concave}
For each $k = 1, 2, \ldots, L$, the density $f_k(x) \propto \frac{1}{\Gamma(x)^J} \ x^{A-1}\ e^{-B_kx}\ \mathds{1}_{\left\{ x > 0 \right\}}$ is log-concave.
\end{lemma}

\begin{proof}
For any $k \in \{1, 2, \ldots, L\}$, define the log density as
\begin{equation} 
\label{log f_k}
    h_k(x) = \log \, f_k(x) = - \log \, C_{f_k} - J \,\log \,\Gamma(x) + (A-1) \, \log \, x - B_kx
\end{equation}
where $C_{f_k}$ denotes the normalizing constant of $f_k$.

Differentiating (\ref{log f_k}) twice with respect to $x$, we have $h_k^{\prime \prime} (x) = -J \psi^\prime(x) - (A-1) /{x^2} $.

For proving concavity of $h_k$, it suffices to show that for all $x>0,\, J \geq 1,\, A \in (0,1)$,
\begin{equation*}
    x^2 \,\psi^\prime(x) > \frac{1-A}{J}
\end{equation*}

Noting $(1-A)/J \in (0,1)$, we shall show that $ x^2 \psi^\prime(x)$ is a strictly increasing function of $x$ in $(0,\infty)$ and $\lim_{x \rightarrow 0^{+}} x^2 \psi^\prime(x) = 1$ which will complete our proof by implying that 
$x^2 \psi^\prime(x) > 1 > {(1-A)}/{J} $.

For $x>0$, let
\begin{equation*}
    g(x) = x^2 \psi^\prime(x)
\end{equation*}
Hence, 
\begin{equation*}
    g^\prime(x) = x \, \{ 2\psi^\prime(x) + x \psi^{\prime\prime}(x)\}
\end{equation*}

For $x \in \bbR^{+}$ and $n \in \bbN$, 
integral representation of a polygamma function, $\psi^{(n)}(x)$ is as follows 
\begin{equation}
\label{integral_polygamma}
    \psi^{(n)}(x) = \frac{d^{n+1}}{dx^{n+1}} \, \log \Gamma(x) = (-1)^{(n+1)} \int_0^\infty \frac{t^n}{1 - e^{-t}} \ e^{-xt}\, dt
\end{equation}
and we have the following recusion formula
\begin{equation}
\label{recursion}
    \psi^{(n-1)}(x+1) = \psi^{(n-1)}(x) + (-1)^{(n-1)}\frac{(n-1)!}{x^n}
\end{equation}

From (\ref{integral_polygamma}) and (\ref{recursion}), it follows that
\begin{equation}
\label{lim_psi}
    \lim_{x \rightarrow 0^{+}} x^n \,\psi^{(n-1)}(x) = (-1)^n \, (n-1)! \ ,
\end{equation}
details of which can be found in \cite{Yang2017SomePO} and \cite{Qi2015CompleteMO}.

Putting $n=2$ in (\ref{lim_psi}), we get
\begin{equation}
\label{lim_g}
    \lim_{x \rightarrow 0^{+}} g(x) = 1
\end{equation}

\vspace{1ex}
Next, for proving strict monotonicity of $g$, we put $n=2$ in (\ref{integral_polygamma}) and integrate by parts to get
\begin{align}
\label{xpsi2}
    \nonumber x\, \psi^{\prime \prime}(x) 
    & = -x \int_0^\infty \frac{t^2}{1 - e^{-t}} \ e^{-xt}\, dt = \int_0^\infty \frac{t^2}{1 - e^{-t}} \ \frac{d}{dt} \left\{e^{-xt}\right\}\, dt \\
    & = - \int_0^\infty \frac{(2e^t - 2 - t)\, t e^t}{(e^t-1)^2}  \ e^{-xt}\ dt
\end{align}

Finally, putting $n=1$ in (\ref{integral_polygamma}) and using (\ref{xpsi2}), we have for all $x>0$,
\begin{align*}
    2 \, \psi^\prime(x) + x\, \psi^{\prime \prime}(x) 
    & = 2 \int_0^\infty \frac{t}{1 - e^{-t}} \ e^{-xt}\, dt - \int_0^\infty \frac{(2e^t - 2 - t)\, t e^t}{(e^t-1)^2}  \ e^{-xt}\ dt \\
    & = \int_0^\infty \frac{t^2 e^t}{(e^t-1)^2}  \ e^{-xt}\ dt \quad > 0
\end{align*}
which thereby implies that $g^\prime(x) > 0$ for all $x>0$ and combining with (\ref{lim_g}) completes our proof.

\end{proof}

\begin{lemma}
\label{lemma:mode-neg-B}
When $B_k<0$, the mode of the density $f_k(x) \propto \frac{1}{\Gamma(x)^J} \ x^{A-1}\ e^{-B_kx}\ \mathds{1}_{\left\{ x > 0 \right\}}$ lies in $(0,\, e^{1-B_k/J})$, for each $k=1, 2, \ldots, L$.
\end{lemma}

\begin{proof}
For $x>0$, consider the function $g(x) = \psi(e^{1+x})-x$. Then $g^\prime (x) = e^{1+x} \, \psi^\prime (e^{1+x}) - 1 >0$, since for all $x>0$,
\begin{align*}
    x\,\psi^\prime(x) & = x \int_0^\infty \frac{t}{1 - e^{-t}} \ e^{-xt}\, dt = - \int_0^\infty \frac{t}{1 - e^{-t}} \, \frac{d}{dt} \{e^{-xt}\}\, dt \\
    & = 1 + \int_0^\infty \frac{e^t(e^t -1 - t)}{(e^t -1)^2}\, e^{-xt}\, dt \quad > \ 1
\end{align*}
\ie $g(x)$ is strictly increasing on $(0, \infty)$ which thereby implies $g(x) > g(0) = \psi(e) >0$ for $x>0$.

Recall the log density defined in (\ref{log f_k}) and note that $h^\prime_k(x) < 0$ is equivalent to 
\begin{equation}
\label{mode_cond}
    x\left\{\psi(x) + \frac{B_k}{J} \right\} > \frac{(A-1)}{J}
\end{equation}
and the right-hand side of (\ref{mode_cond}) is negative, as $A \in (0,1)$ and $J \in \mathbb{N}$.

Plugging $x = e^{1 - B_k/J}$ in (\ref{mode_cond}), the left-hand side boils down to $e^{1 - B_k/J}g(- B_k/J)>0$, thereby implying that $h^\prime_k(e^{1 - B_k/J})<0$. Consequently, the log-concavity of $f_k$ ensures that its mode lies in $(0,e^{1-B_k/J})$.
\end{proof}

\section{Sensitivity analyses}
\label{sensitivity}
In this section, we outline how sensitive the performance of our algorithm is against different choices of the truncation level $L$ of the underlying random measures and the hyperparameters $(\gamma, b_0)$ which specify the gamma prior on $\alpha_0$. The data is generated in the same way as outlined in \S \ref{simulation}.

\subsection{Choice of truncation level}
\begin{figure}[htp]
    \centering
    \begin{subfigure}[b]{0.34\textwidth}
         \centering
         \includegraphics[width=\textwidth]{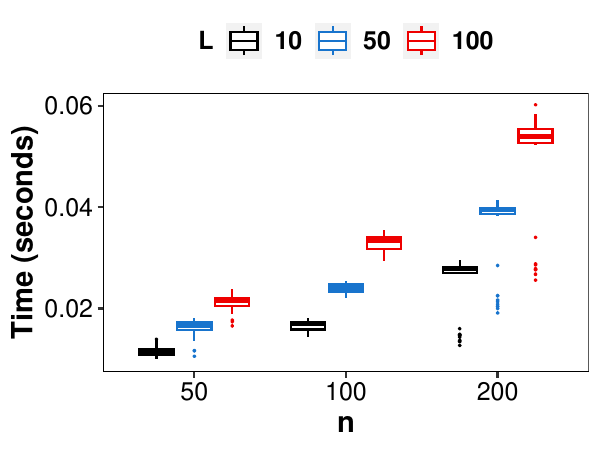}
         \caption{}
     \end{subfigure}
     \hspace{1.5cm}
     \begin{subfigure}[b]{0.34\textwidth}
         \centering
         \includegraphics[width=\textwidth]{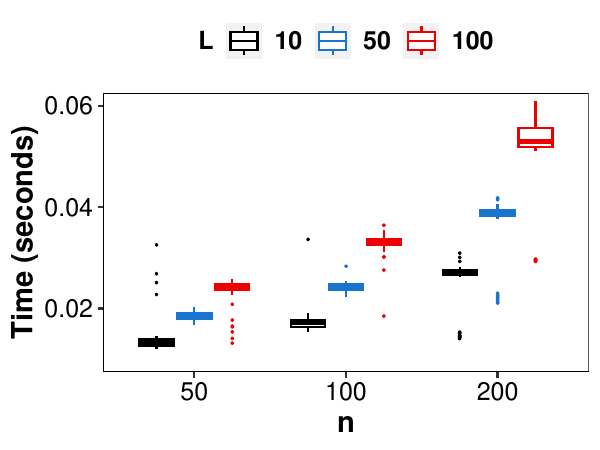}
         \caption{}
     \end{subfigure}
    \caption{Average computation time (in seconds) per MCMC iteration, when true means of the Gaussian mixture are (a) $\uphi^{0} = (-3,-1,1,3)$ and (b) $\uphi^{0} = (-6,-2,2,6)$, across different choices of the truncation level $L$ under the BGS algorithm. Boxplots show variation across 50 simulation replicates.}
    \label{fig:Time_L}
\end{figure}

\begin{figure}[htp]
     \centering
     \begin{subfigure}[b]{0.85\textwidth}
         \centering
         \includegraphics[width=\textwidth]{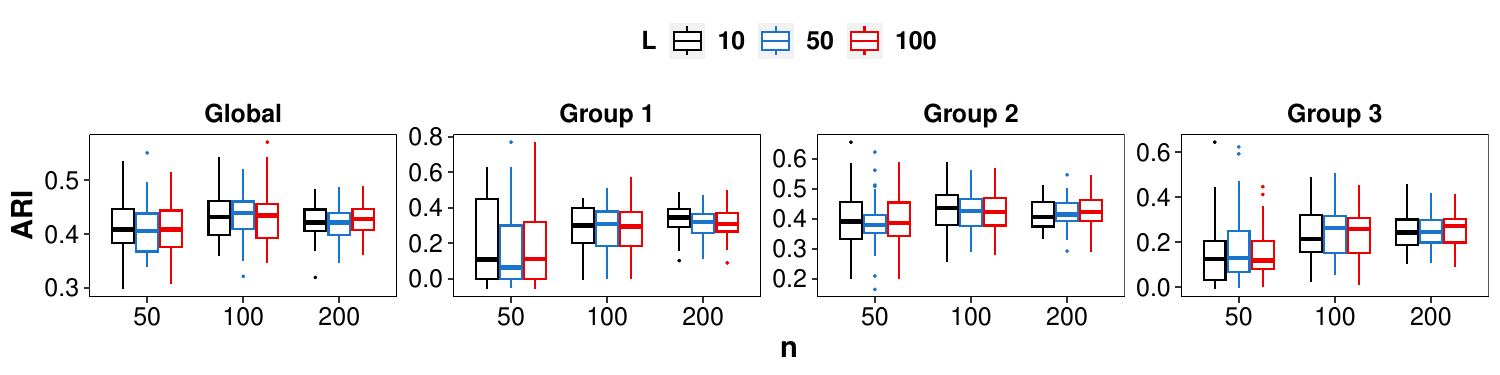}
         \caption{}
     \end{subfigure}
     \begin{subfigure}[b]{0.85\textwidth}
         \centering
         \includegraphics[width=\textwidth]{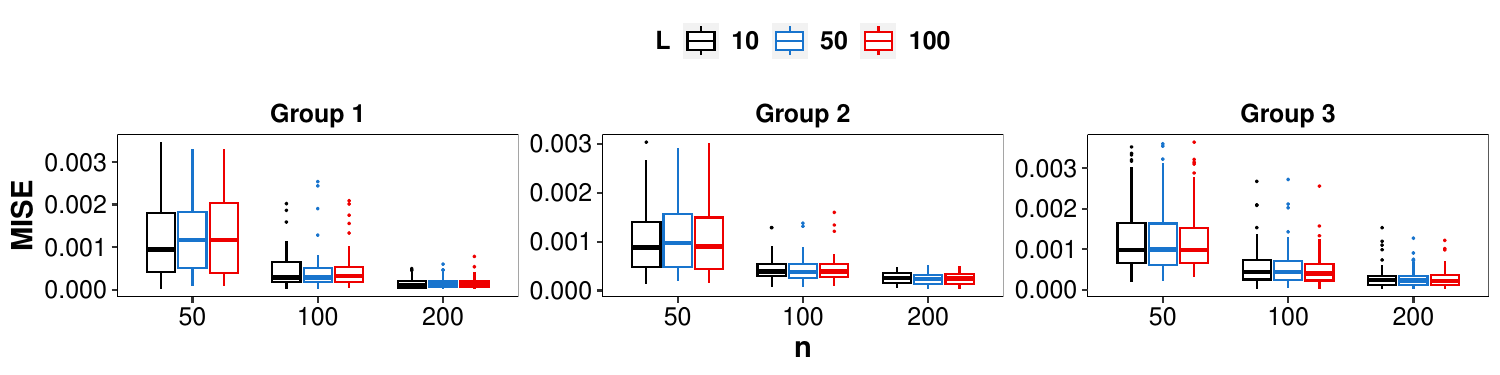}
         \caption{}
     \end{subfigure}
     \begin{subfigure}[b]{0.85\textwidth}
         \centering
         \includegraphics[width=\textwidth]{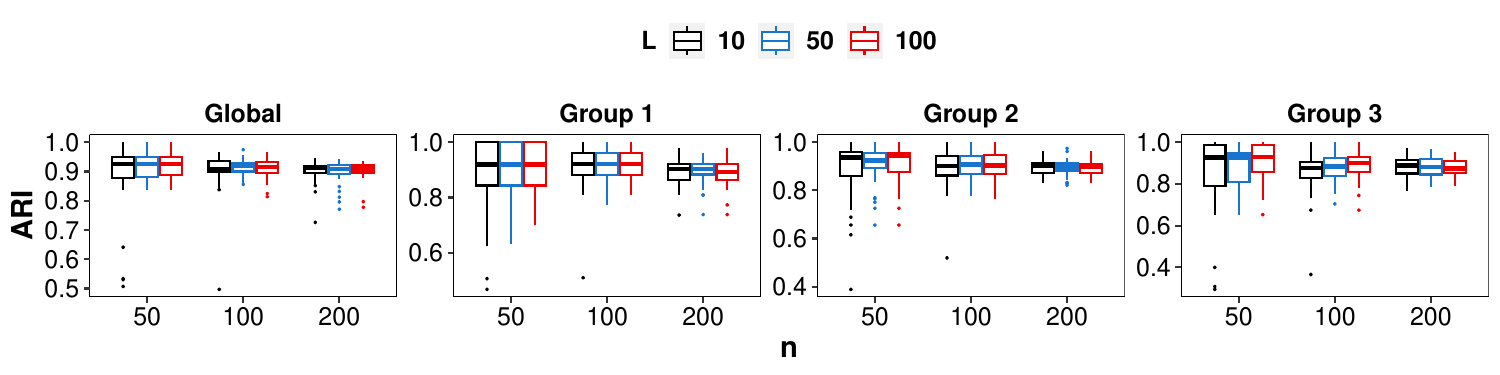}
         \caption{}
     \end{subfigure}
     \begin{subfigure}[b]{0.85\textwidth}
         \centering
         \includegraphics[width=\textwidth]{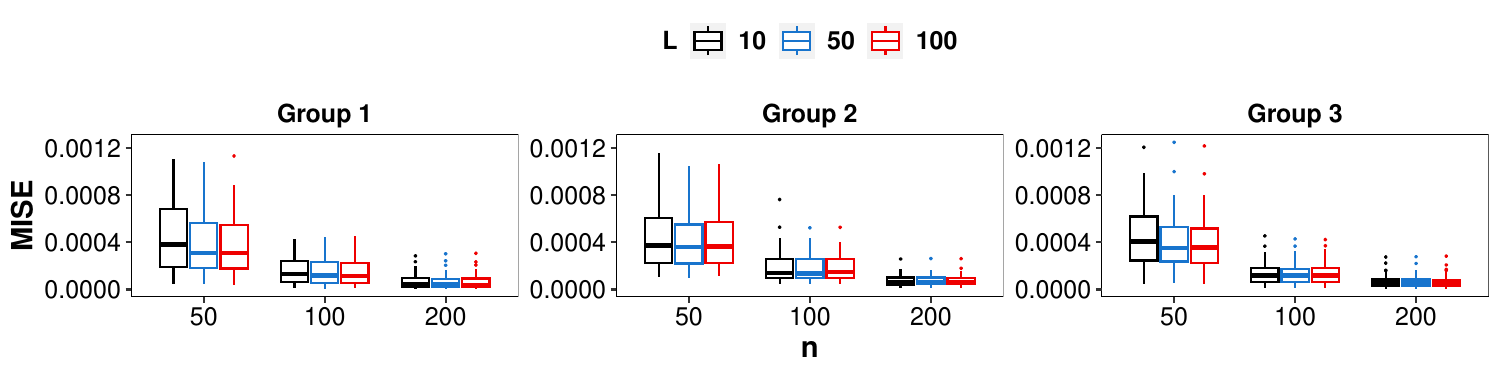}
         \caption{}
     \end{subfigure}
    \caption{Adjusted Rand indices -- (a),(c) of the estimated cluster labels and mean integrated squared error -- (b),(d) of the estimated densities, when true means of the Gaussian mixture are $\uphi^{0} = (-3,-1,1,3)$ --(a),(b) and $\uphi^{0} = (-6,-2,2,6)$ -- (c),(d), across different choices of the truncation level $L$ under the BGS algorithm. Boxplots show variation across 50 simulation replicates.}
    \label{fig:stat_acc_L}
\end{figure}

\begin{figure}[htp]
     \centering
     \begin{subfigure}[b]{0.85\textwidth}
         \centering
         \includegraphics[width=\textwidth]{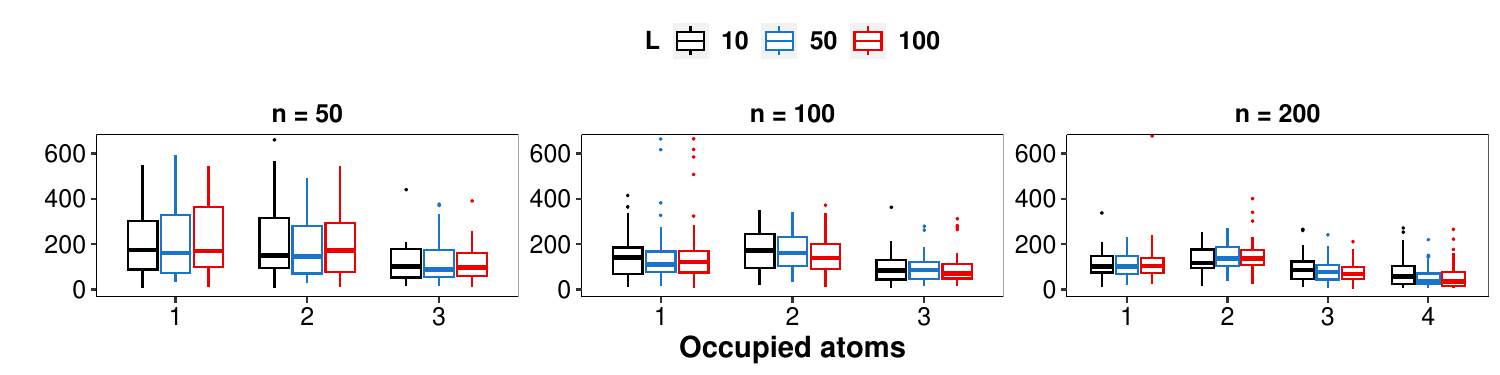}
         \caption{}
     \end{subfigure}
     \begin{subfigure}[b]{0.85\textwidth}
         \centering
         \includegraphics[width=\textwidth]{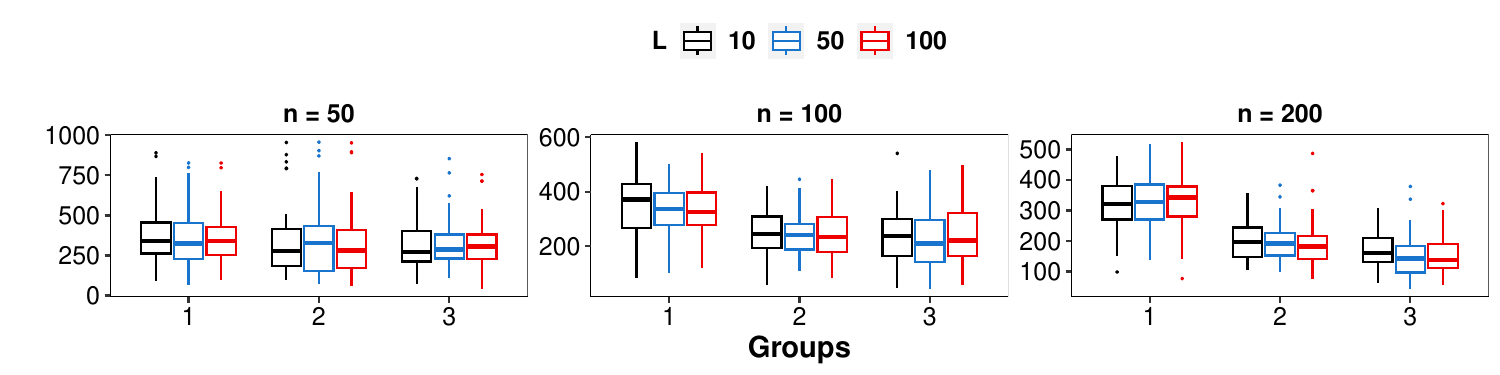}
         \caption{}
     \end{subfigure}
     \begin{subfigure}[b]{0.85\textwidth}
         \centering
         \includegraphics[width=\textwidth]{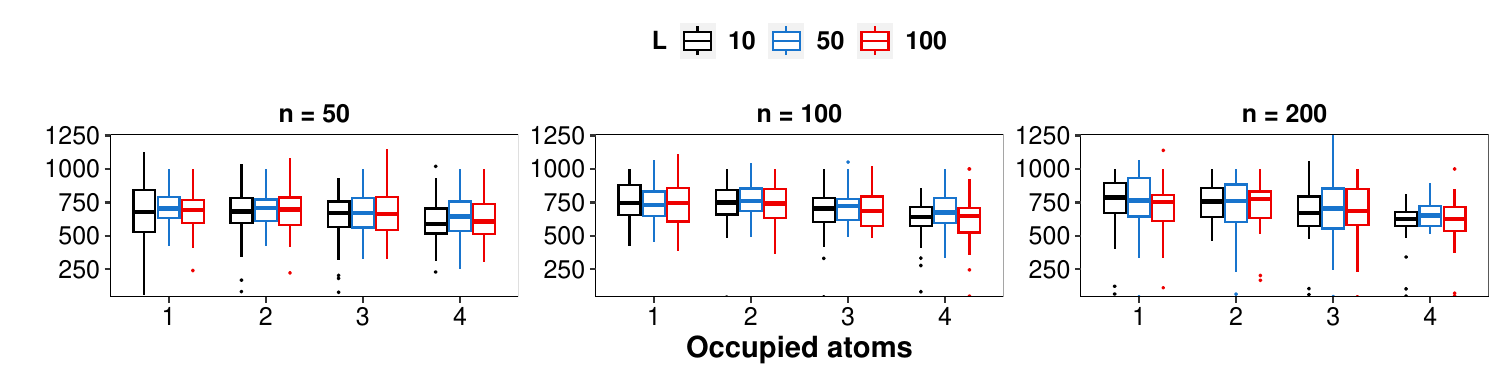}
         \caption{}
     \end{subfigure}
     \begin{subfigure}[b]{0.85\textwidth}
         \centering
         \includegraphics[width=\textwidth]{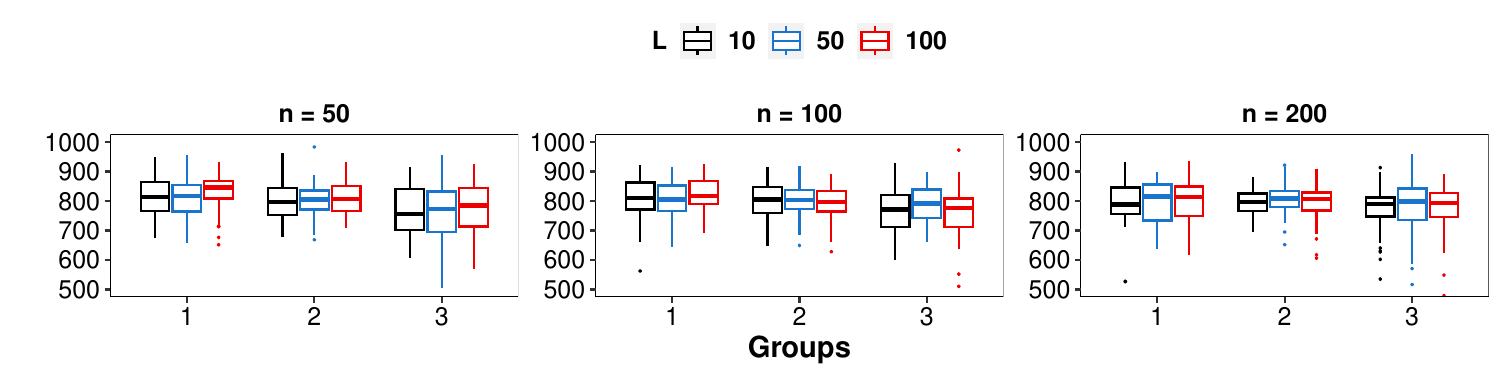}
         \caption{}
     \end{subfigure}
    \caption{Effective sample sizes of occupied atoms -- (a),(c) and estimated densities -- (b),(d), when true means of the Gaussian mixture are $\uphi^{0} = (-3,-1,1,3)$ -- (a),(b) and $\uphi^{0} = (-6,-2,2,6)$ -- (c), (d), across different choices of the truncation level $L$ under the BGS algorithm. Boxplots show variation across 50 simulation replicates.}
    \label{fig:Mixing_L}
\end{figure}

We evaluated our algorithm's performance by altering the truncation level $L$ over $\{10, 50, 100\}$, while maintaining fixed hyperparameters at $\gamma = 1$ and $b_0 = 0.1$, as specified in all simulations in \S\ref{simulation}. Figure \ref{fig:stat_acc_L} illustrates the statistical performance of our algorithm, while Figure \ref{fig:Mixing_L} depicts the mixing behaviour under both overlapping and well-separated designs. The performance of our algorithm remains robust across different choices of $L$. Figure \ref{fig:Time_L} shows boxplots of the average computation time per MCMC iteration with increasing $L$, for each sample size. Scalability of the sampler decreases with increase in $L$, as one would naturally expect.

\subsection{Choice of hyperparameters}
\begin{figure}[htp]
     \centering
     \begin{subfigure}[b]{0.85\textwidth}
         \centering
         \includegraphics[width=\textwidth]{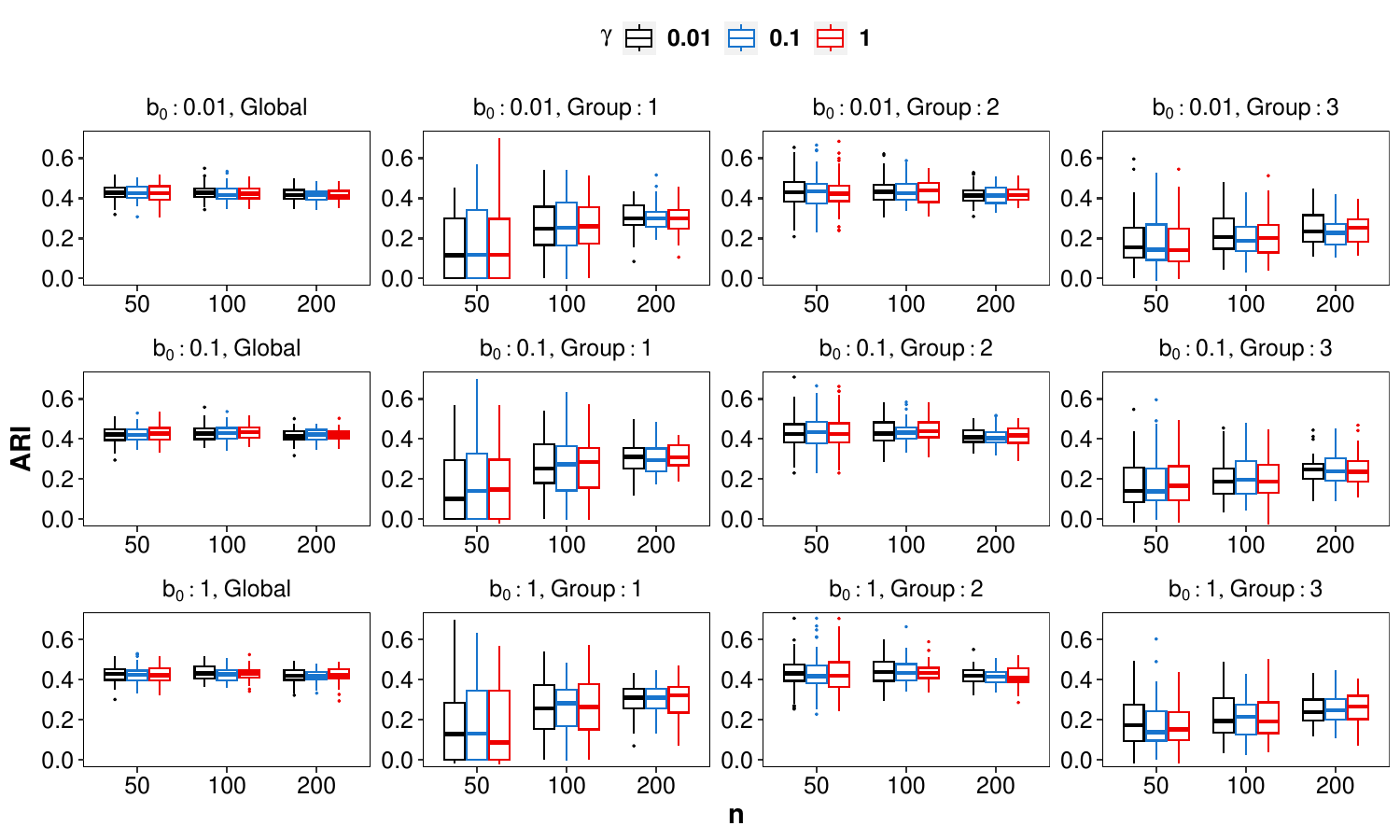}
         \caption{}
     \end{subfigure}
     \begin{subfigure}[b]{0.85\textwidth}
         \centering
         \includegraphics[width=\textwidth]{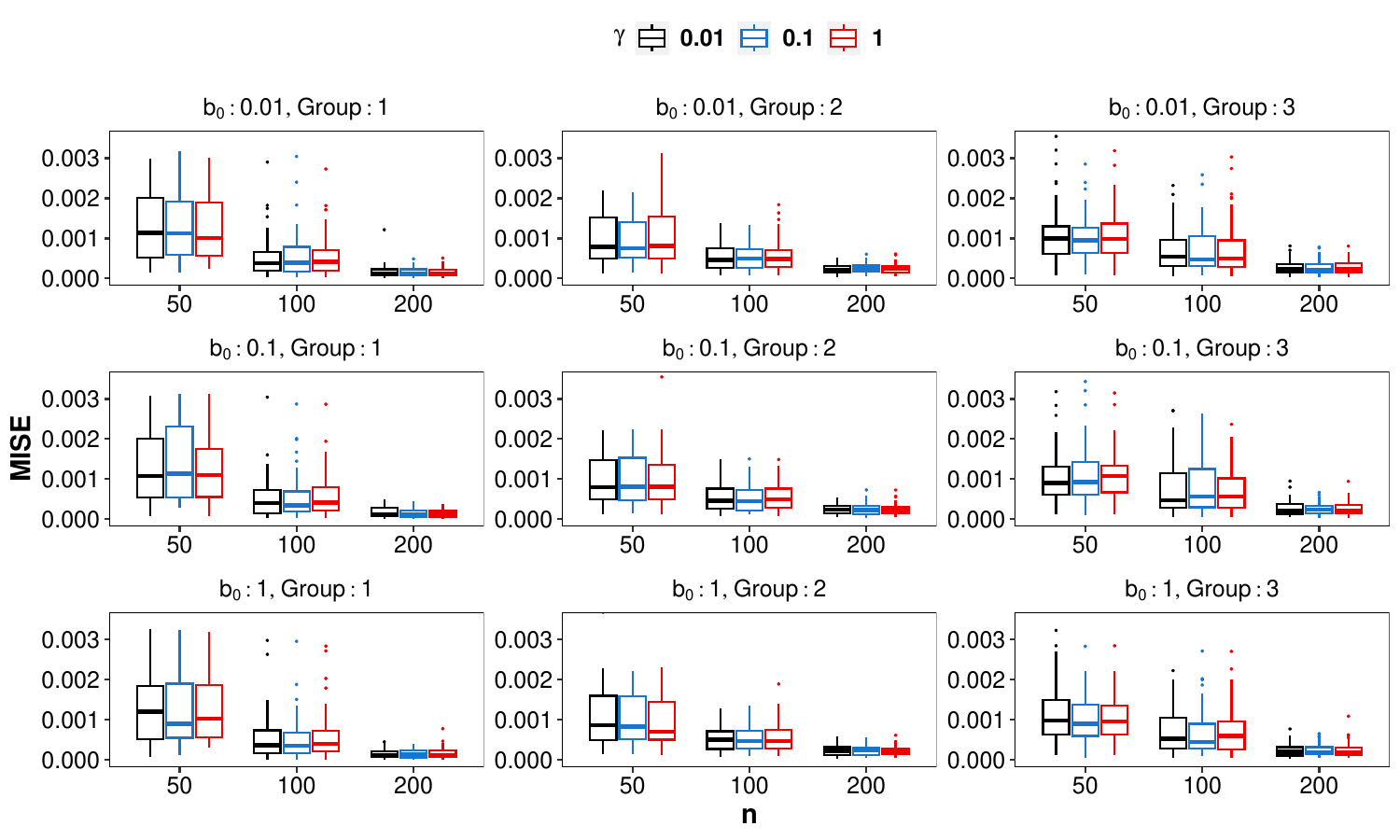}
         \caption{}
     \end{subfigure}
    \caption{(a) Adjusted Rand indices of the estimated cluster labels and (b) mean integrated squared error of the estimated densities, when true means of the Gaussian mixture are $\uphi^{0} = (-3,-1,1,3)$, across different choices of hyperparameters $(\gamma, b_0)$ under the BGS algorithm. Boxplots show variation across 50 simulation replicates.}
    \label{fig:stat_acc_gamma_b0_1}
\end{figure}

\begin{figure}[htp]
     \centering
     \begin{subfigure}[b]{0.85\textwidth}
         \centering         \includegraphics[width=\textwidth]{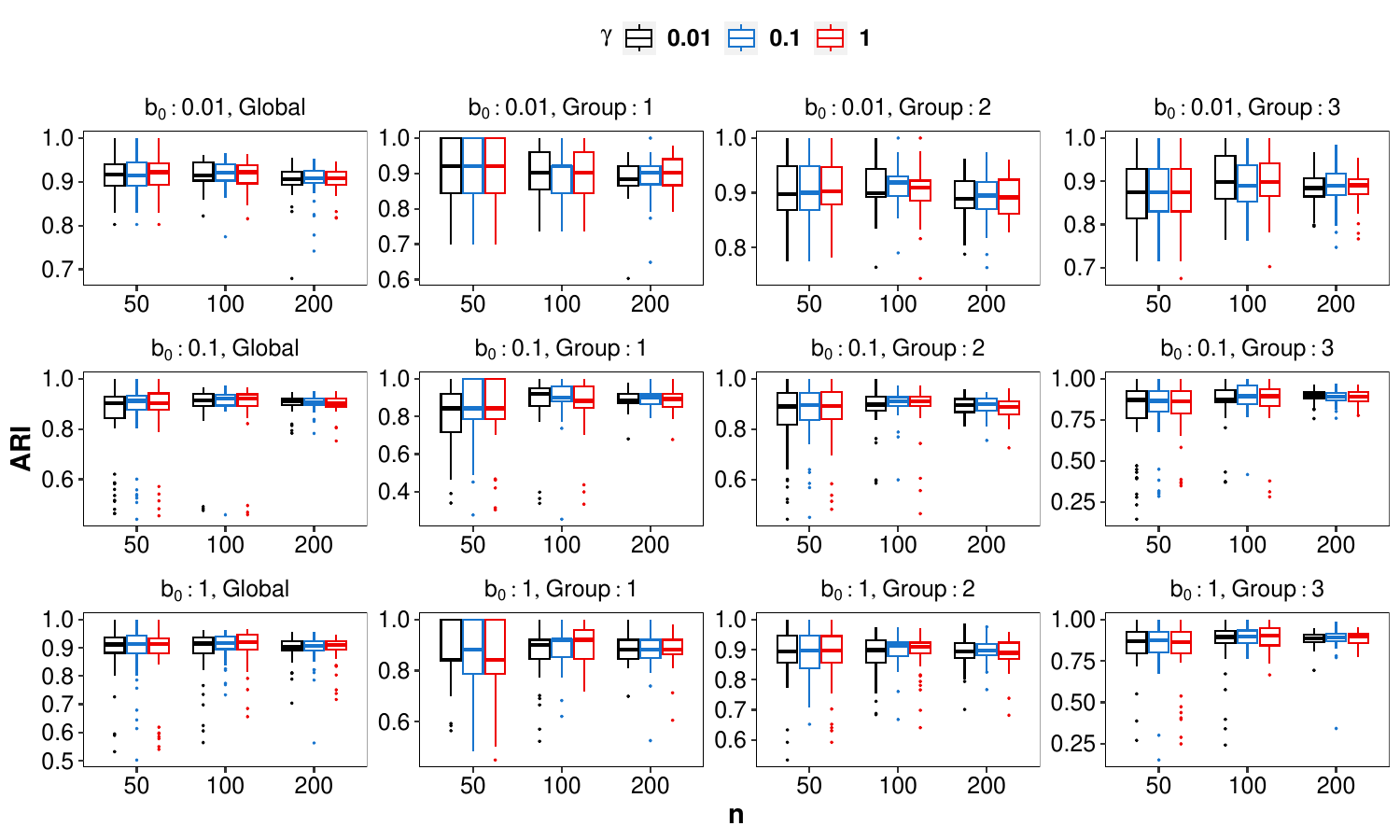}
         \caption{}
     \end{subfigure}
     \begin{subfigure}[b]{0.85\textwidth}
         \centering
         \includegraphics[width=\textwidth]{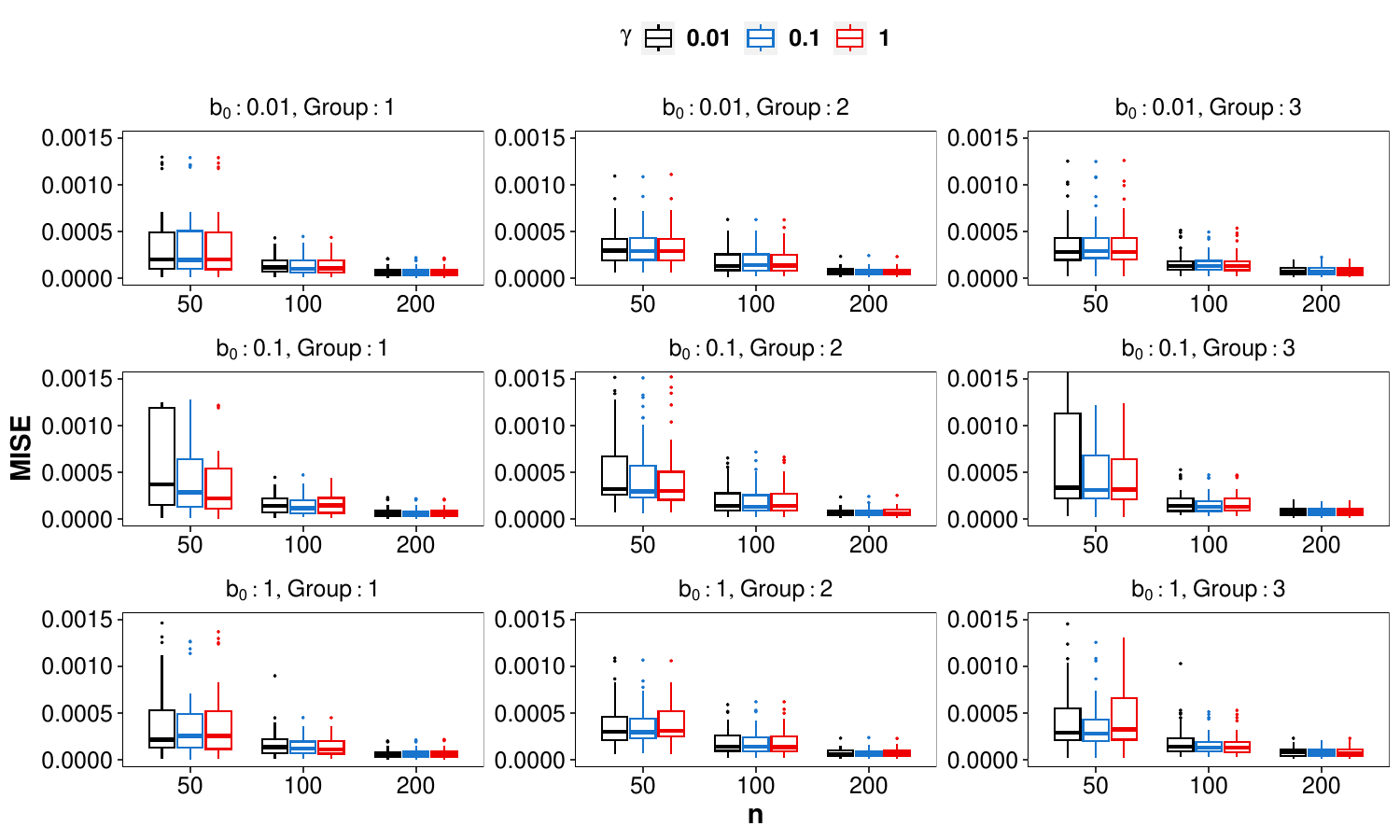}
         \caption{}
     \end{subfigure}
    \caption{(a) Adjusted Rand indices of the estimated cluster labels and (b) mean integrated squared error of the estimated densities, when true means of the Gaussian mixture are $\uphi^{0} = (-6,-2,2,6)$, across different choices of hyperparameters $(\gamma, b_0)$ under the BGS algorithm. Boxplots show variation across 50 simulation replicates.}
    \label{fig:stat_acc_gamma_b0_2}
\end{figure}

\begin{figure}[htp]
     \centering
     \begin{subfigure}[b]{0.85\textwidth}
         \centering
         \includegraphics[width=\textwidth]{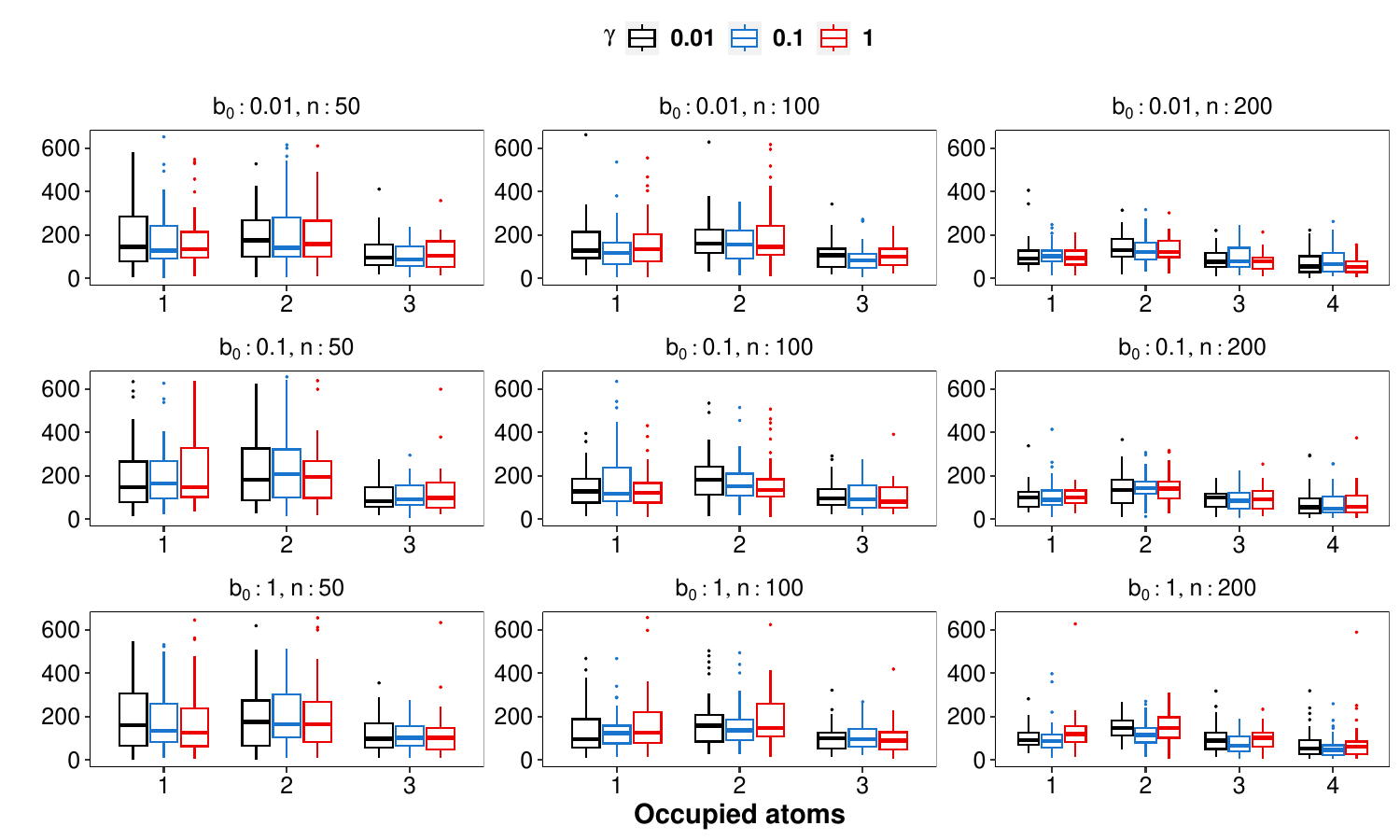}
         \caption{}
     \end{subfigure}
     \begin{subfigure}[b]{0.85\textwidth}
         \centering
         \includegraphics[width=\textwidth]{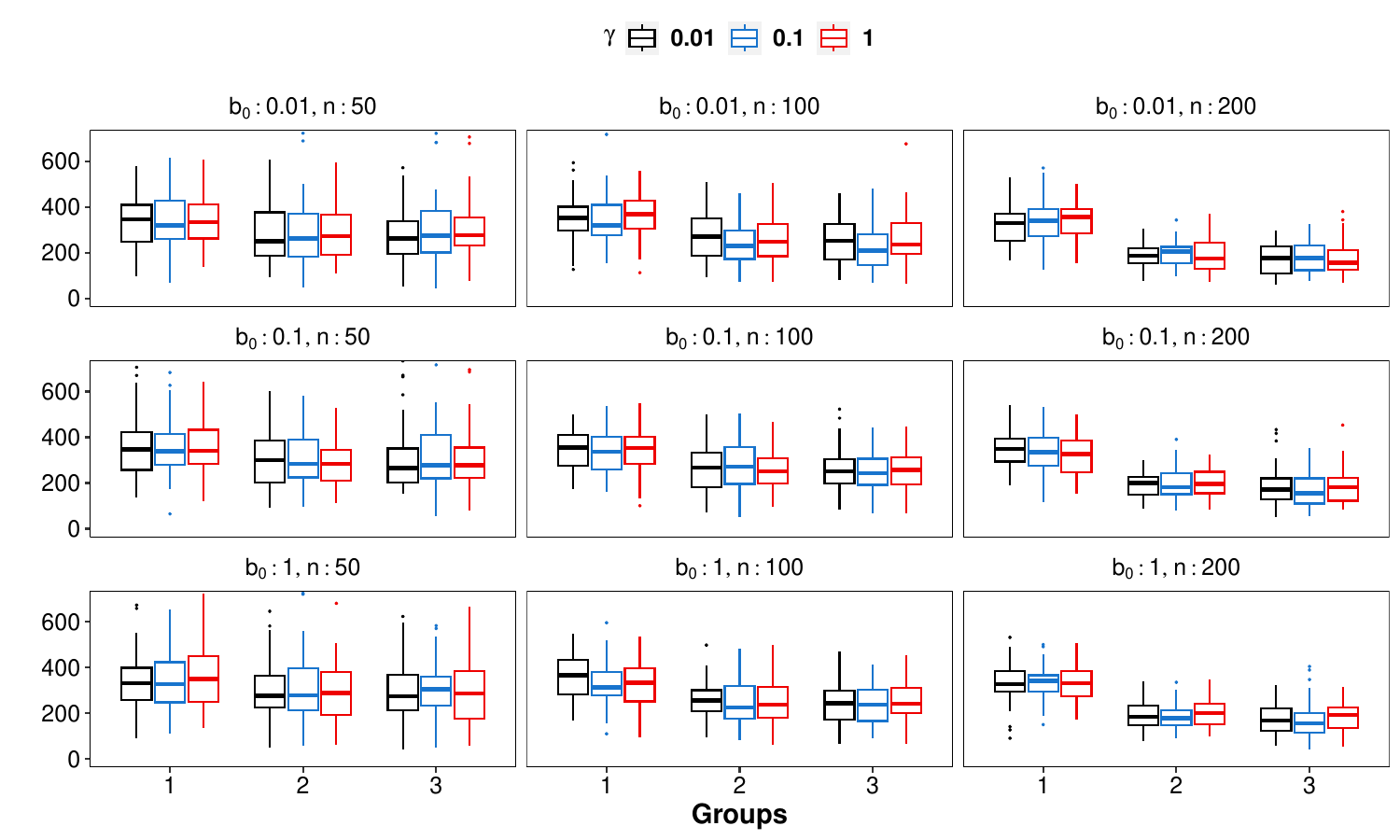}
         \caption{}
     \end{subfigure}
    \caption{Effective sample sizes of (a) occupied atoms and (b) estimated densities, when true means of the Gaussian mixture are $\uphi^{0} = (-3,-1,1,3)$, across different choices of hyperparameters $(\gamma, b_0)$ under the BGS algorithm. Boxplots show variation across 50 simulation replicates.}
    \label{fig:Mixing_gamma_b0_1}
\end{figure}

\begin{figure}[htp]
     \centering
     \begin{subfigure}[b]{0.85\textwidth}
         \centering
         \includegraphics[width=\textwidth]{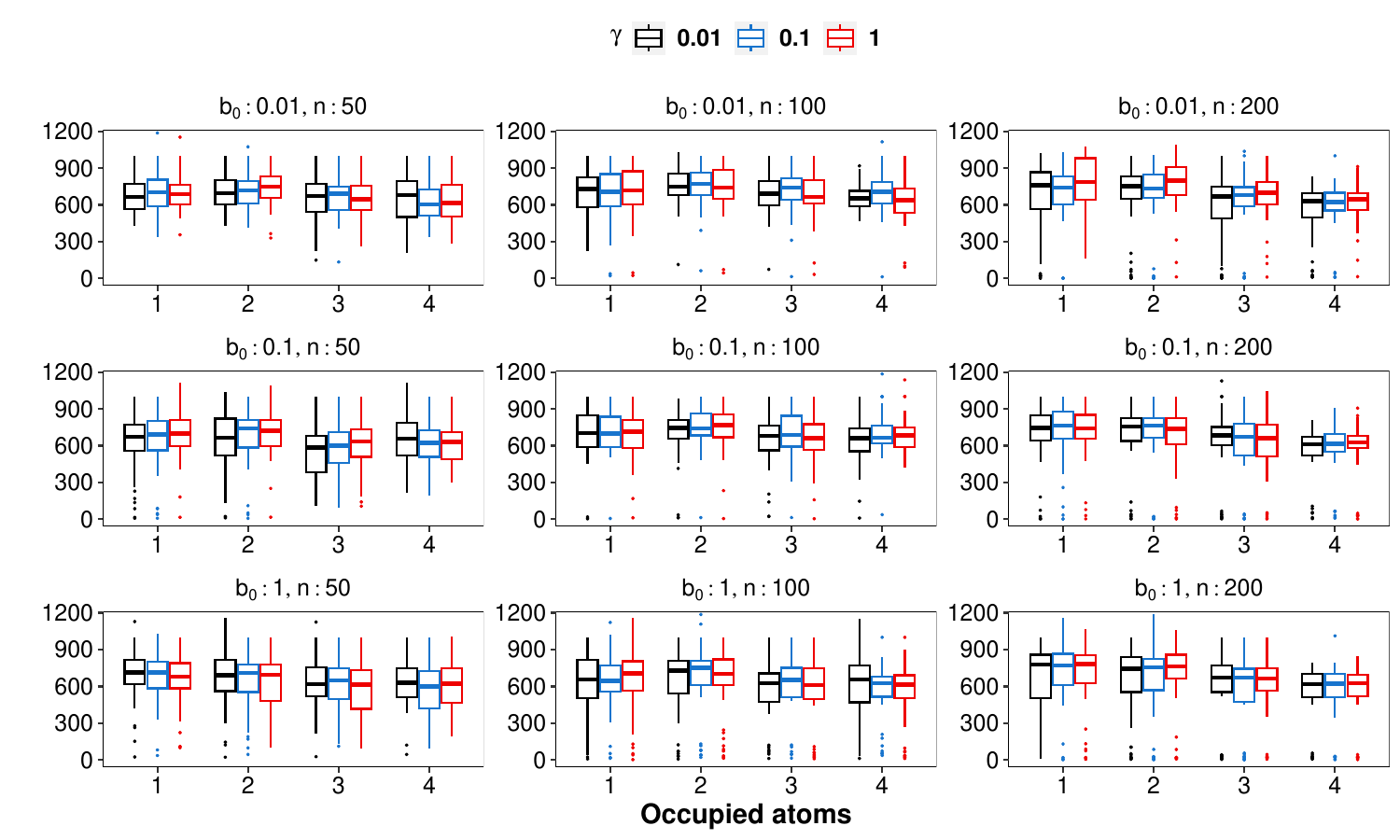}
         \caption{}
     \end{subfigure}
     \begin{subfigure}[b]{0.85\textwidth}
         \centering
         \includegraphics[width=\textwidth]{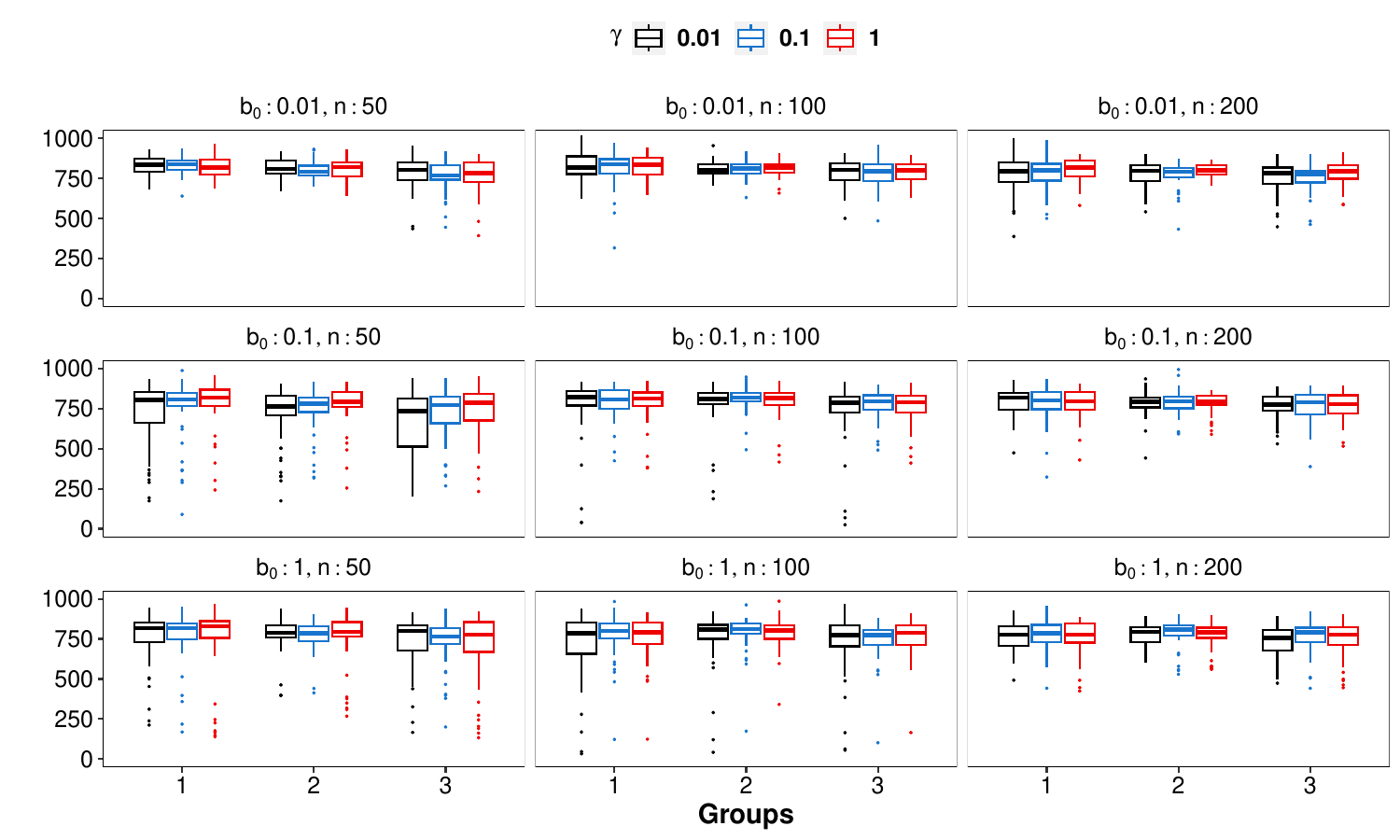}
         \caption{}
     \end{subfigure}
    \caption{Effective sample sizes of (a) occupied atoms and (b) estimated densities, when true means of the Gaussian mixture are $\uphi^{0} = (-6,-2,2,6)$, across different choices of hyperparameters $(\gamma, b_0)$ under the BGS algorithm. Boxplots show variation across 50 simulation replicates.}
    \label{fig:Mixing_gamma_b0_2}
\end{figure}

We implemented our algorithm by varying both $\gamma$ and $b_0$ over $\{0.01, 0.1, 1\}$, while keeping the truncation level fixed at $L=10$. Figures \ref{fig:stat_acc_gamma_b0_1} and \ref{fig:stat_acc_gamma_b0_2} depict the statistical performance of our algorithm, while Figures \ref{fig:Mixing_gamma_b0_1} and \ref{fig:Mixing_gamma_b0_2} show the mixing behaviour under overlapping and well-separated designs respectively. Our sampler demonstrates robust performance across various hyperparameter choices.

\newpage

\section{Additional plots}
In the following, we provide additional plots that illustrate the computational efficiency and estimation performance of the competing algorithms described in \S\ref{simulation} of the main document.

\subsection{Computational efficiency of competing algorithms}
\label{addplots_time}
\begin{figure}[htp]
    \centering
    \begin{subfigure}[b]{0.33\textwidth}
         \centering
         \includegraphics[width=\textwidth]{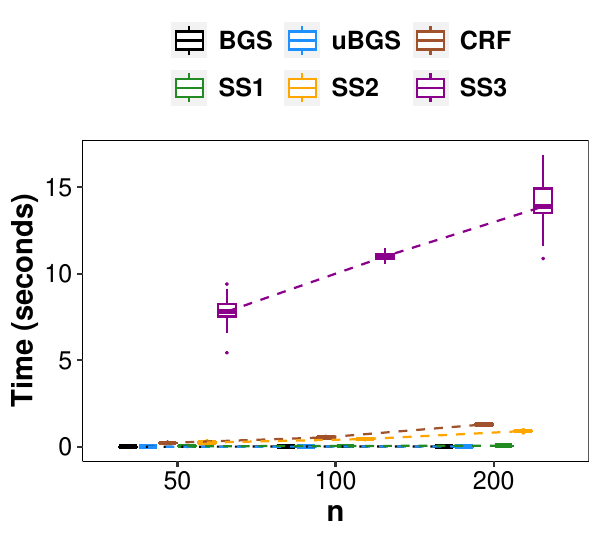}
         \caption{}
     \end{subfigure}
     \hspace{1.5cm}
     \begin{subfigure}[b]{0.33\textwidth}
         \centering
         \includegraphics[width=\textwidth]{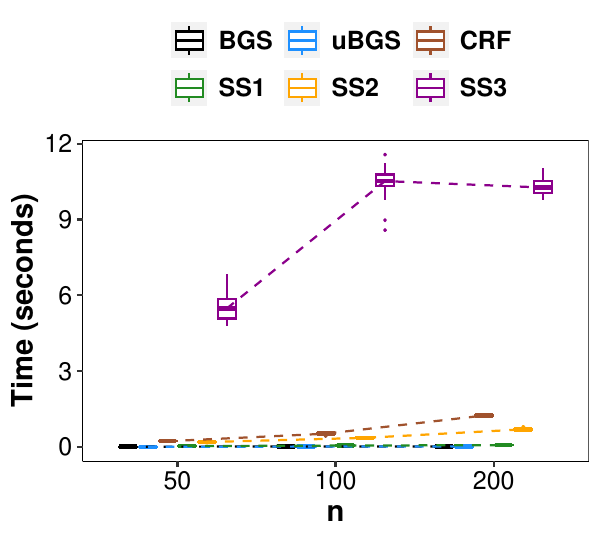}
         \caption{}
     \end{subfigure}
    \caption{Average computation time (in seconds) per MCMC iteration, when true means of the Gaussian mixture are (a) $\uphi^{0} = (-3,-1,1,3)$ and (b) $\uphi^{0} = (-6,-2,2,6)$. SS1, SS2, SS3 refer to the slice samplers with $\alpha_0$ chosen as 0.1, 1, 10 respectively. Boxplots show variation across 50 simulation replicates, along with trajectories of the median time.}
    \label{fig:Time_supp}
\end{figure}

Figure \ref{fig:Time_supp} shows boxplots of the average computation time per MCMC iteration for the competing samplers across 50 simulation replicates with increasing $n$.
The scalability of BGS (comparable to that of uBGS) is ensured by its blocked parameter updates which factorize over the observations in each group, unlike the one-at-a-time updates of CRF that makes it suffer heavily when $n$ is large.   
The computation time for SS is seen to exhibit an increase when a higher value is assigned to $\alpha_0$.

\vspace{0.3cm}

\subsection{Estimation performance of competing algorithms}
\label{addplots_estimation}
We present plots that depict the performance in clustering and density estimation, of BGS, uBGS, CRF and the three implementations of SS in a single simulation replicate considering a sample of size $n = 200$. 

\begin{figure}[htp]
    \centering
    \begin{subfigure}[b]{0.6\textwidth}
         \centering
         \includegraphics[width=\textwidth]{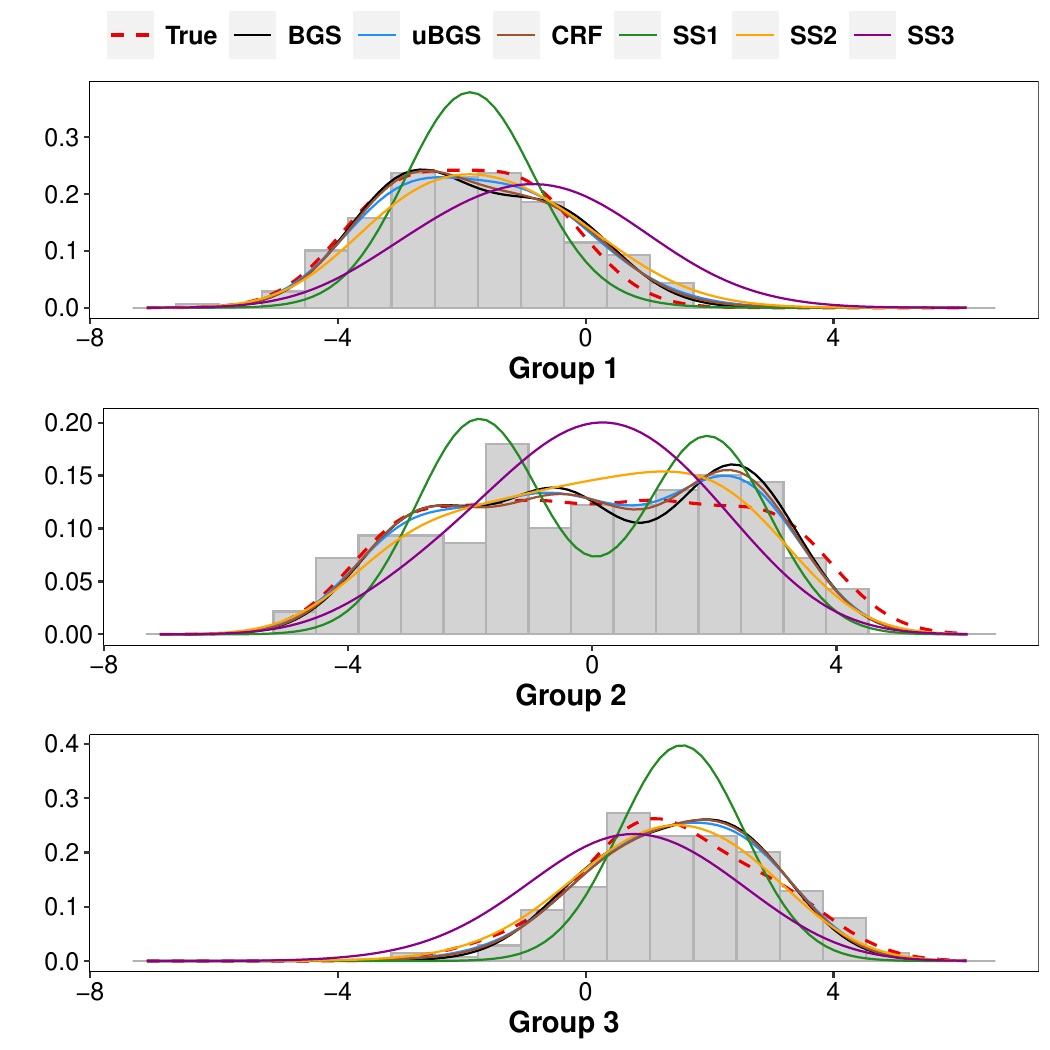}
         \caption{\vspace{2ex}}
     \end{subfigure}
     \begin{subfigure}[b]{0.6\textwidth}
         \centering
         \includegraphics[width=\textwidth]{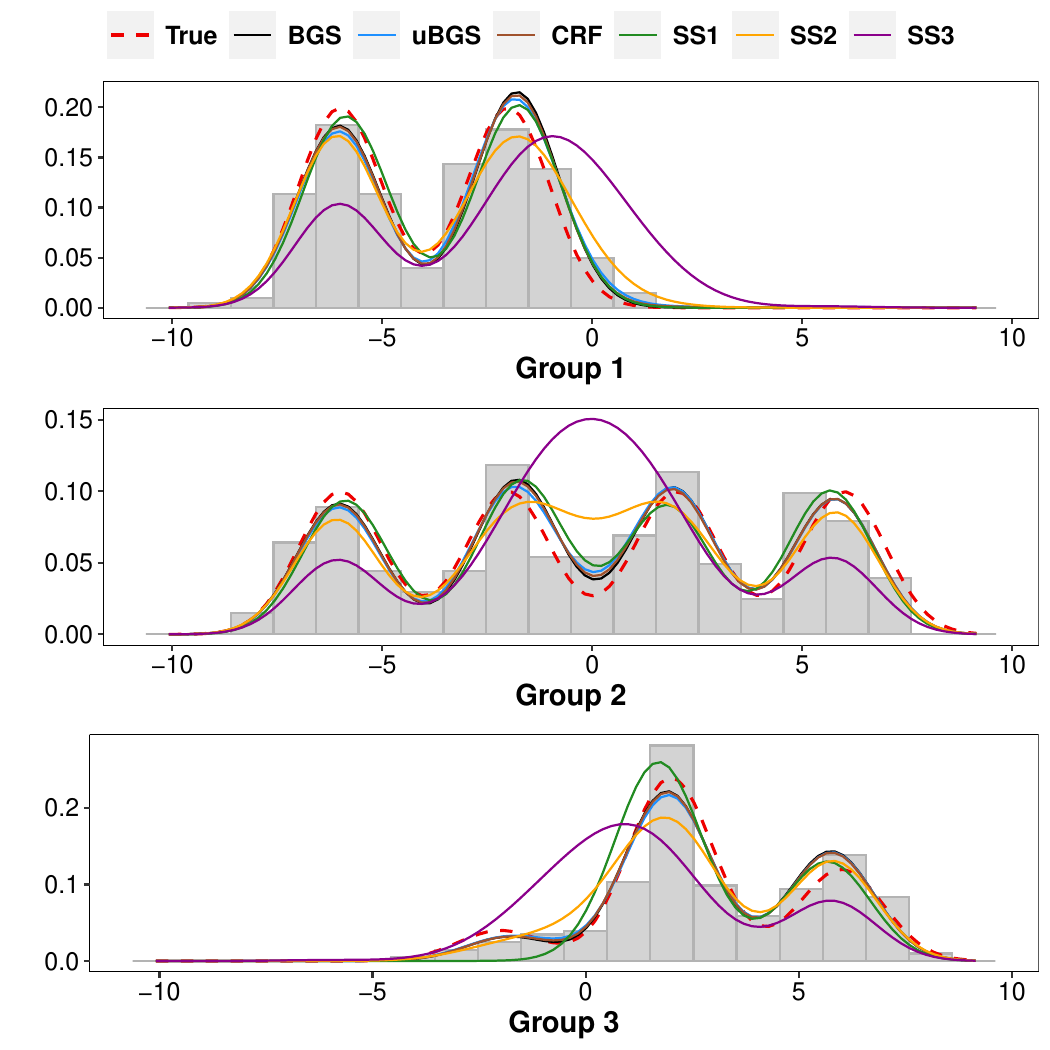}
         \caption{}
     \end{subfigure}
    \caption{True and estimated densities overlaid on the histogram of the observed data, when true means of the Gaussian mixture are (a) $\uphi^{0} = (-3, -1, 1, 3)$ and (b) $\uphi^{0} = (-6, -2, 2, 6)$. SS1, SS2, SS3 refer to the slice samplers with $\alpha_0$ chosen as 0.1, 1, 10 respectively.}
    \label{fig:densityplots}
\end{figure}
\begin{figure}[htp]
    \centering
    \includegraphics[width=13cm]{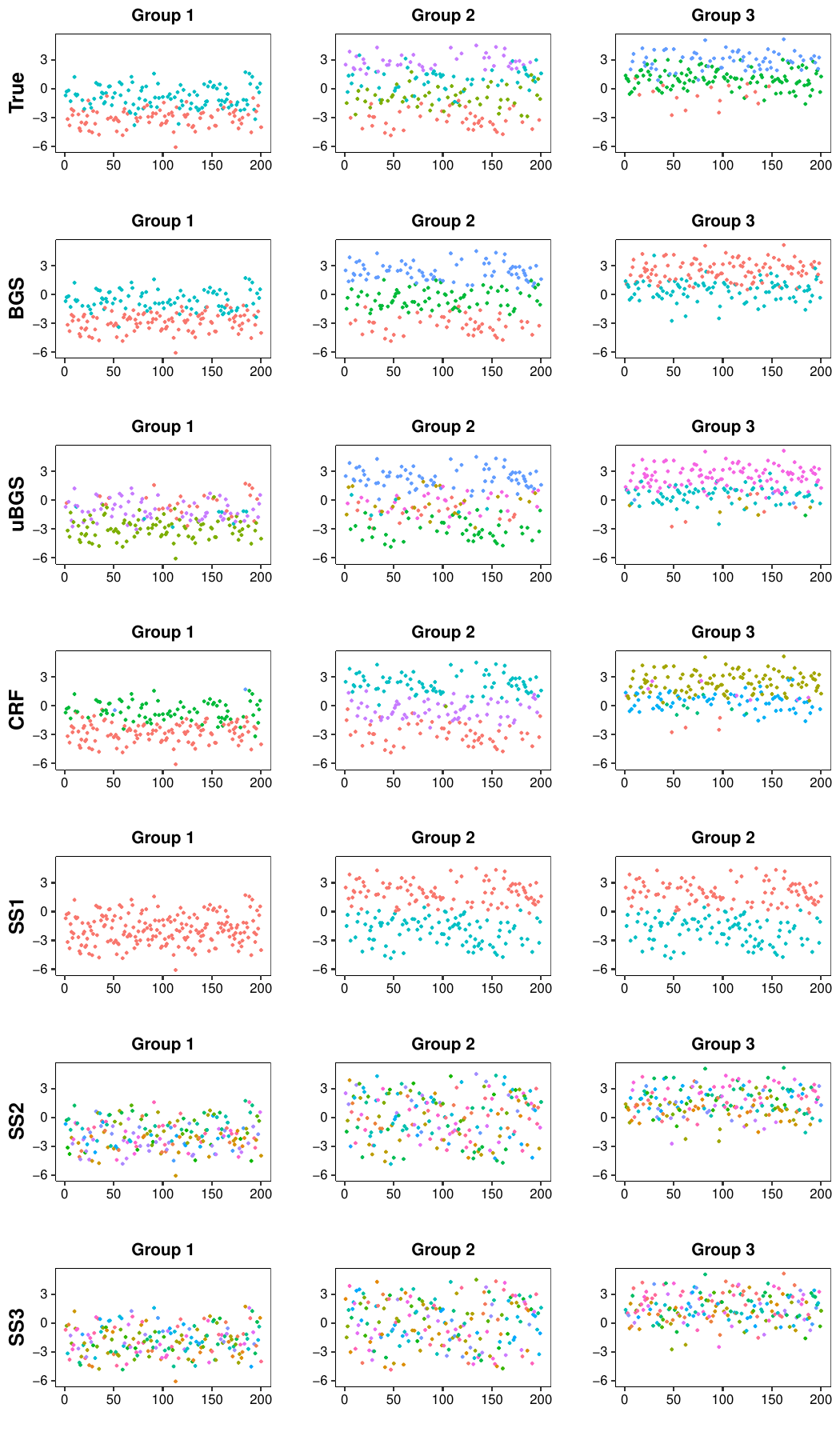}
    \caption{True and estimated cluster labels for a sample of size $n = 200$, when true means of the Gaussian mixture are $\uphi^{0} = (-3, -1, 1, 3)$. SS1, SS2, SS3 refer to the slice samplers with $\alpha_0$ chosen as 0.1, 1, 10 respectively.}
    \label{fig:clusters-(i)}
\end{figure}
\begin{figure}[htp]
    \centering
    \includegraphics[width=13cm]{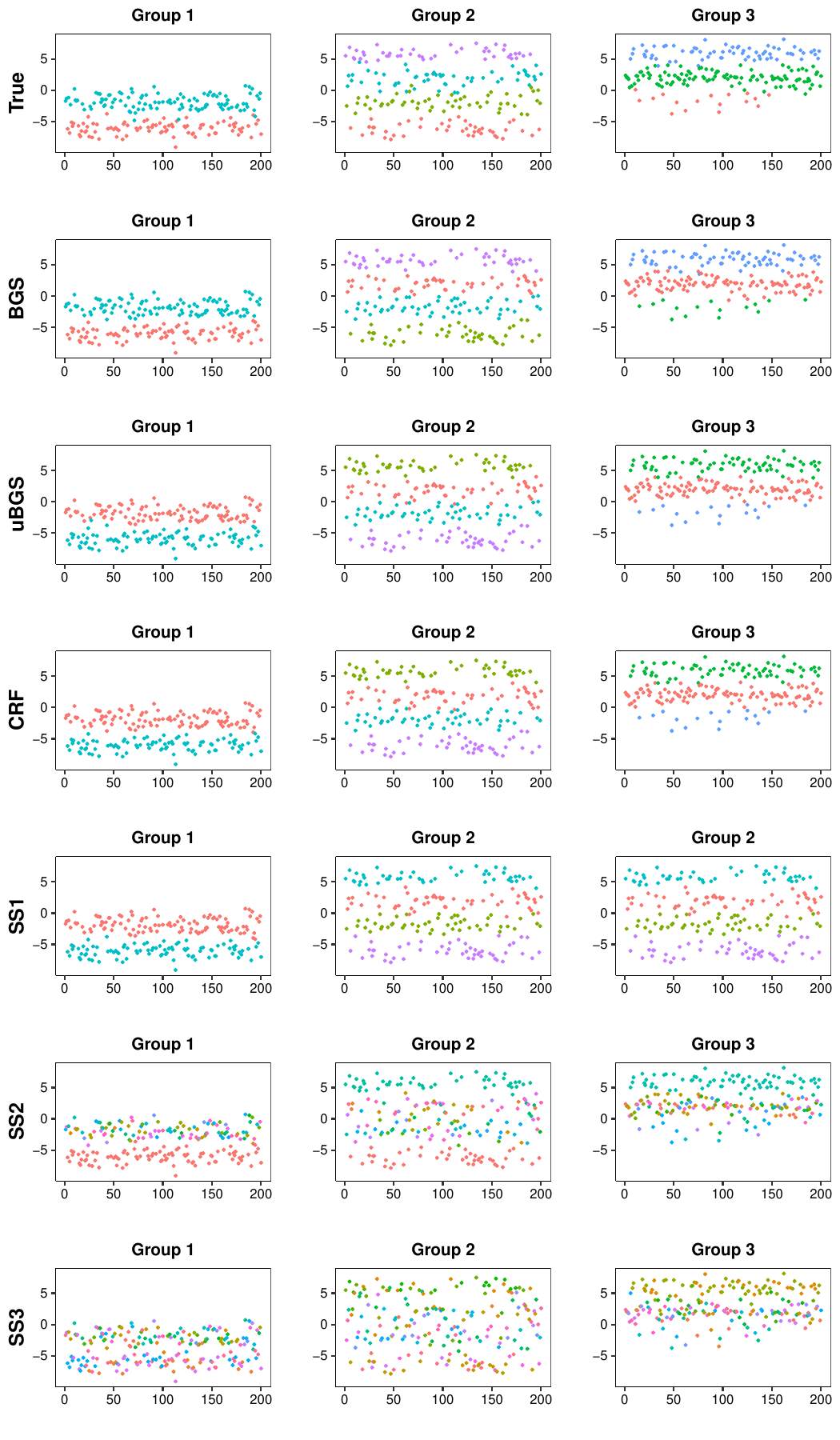}
    \caption{True and estimated cluster labels for a sample of size $n = 200$, when true means of the Gaussian mixture are $\uphi^{0} = (-6, -2, 2, 6)$. SS1, SS2, SS3 refer to the slice samplers with $\alpha_0$ chosen as 0.1, 1, 10 respectively.}
    \label{fig:clusters-(ii)}
\end{figure}

Figure \ref{fig:densityplots} shows the estimated densities of the the three groups along with the true density, overlaid on the histogram of the observed data.
Figures \ref{fig:clusters-(i)} and \ref{fig:clusters-(ii)} show the cluster labels estimated by the algorithms for all the groups along with the true labels under overlapping and well separated designs respectively. 

\end{document}